\documentclass[useAMS,usenatbib]{mn2e}
\usepackage{graphicx}
\usepackage{amssymb, amsmath}
\usepackage{times}
\usepackage{color}
\usepackage{lscape}
\usepackage[section]{placeins}
\usepackage{epstopdf}
\usepackage{epsf}
\usepackage{ulem}

\newcommand{\Om}{\Omega_m}
\newcommand{\Ob}{\Omega_b}
\newcommand{\LCDM}{\rm {\Lambda CDM}}
\newcommand{\OL}{\Omega_\Lambda}
\newcommand{\Ok}{\Omega_K}

\newcommand{\DA}{D\!_A(z)}
\newcommand{\hz}{H(z)}

\newcommand{\Hz}{H\!_z\,}

\newcommand{\DV}{D_V}
\newcommand{\sDV}{\sigma_{D_V}}
\newcommand{\hMpc}{h^{-1}{\rm\;Mpc}}

\newcommand{\trihGpc}{h^{-3}{\rm\;Gpc^3}}
\newcommand{\trihMpc}{h^{-3}{\rm\;Mpc^{3}}}
\newcommand{\itrihMpc}{h^{3}{\rm\;Mpc^{-3}}}
\newcommand{\ihMpc}{h{\rm\;Mpc^{-1}}}
\newcommand{\Mpc}{{\rm\;Mpc}}
\newcommand{\Gpc}{{\rm\;Gpc}}

\newcommand{\sig}{\sigma}

\newcommand{\Sigxy}{\Sigma_{\rm xy}}
\newcommand{\Sigz}{\Sigma_{z}}
\newcommand{\Sigxyo}{\Sigma_{\rm xy,0}}
\newcommand{\Sigzo}{\Sigma_{z,0}}

\newcommand{\Signleff}{\Sigma_{\rm nl,eff}}

\newcommand{\SigR}{\Sigma_{R}}
\newcommand{\SigRo}{\Sigma_{R,0}}

\newcommand{\Plin}{P_{\rm lin}}

\newcommand{\rhoow}{\rho_{0,w}} 
\newcommand{\rhorw}{\rho_{r,w}} 

\newcommand{\hdelta}{\hat{\delta}}
\newcommand{\kmax}{k_{\rm max}}
\newcommand{\mumin}{\mu_{\rm min}}
\newcommand{\kperp}{k_\perp}
\newcommand{\kpar}{k_\parallel}

\newcommand{\kvec}{\vec{k}}

\newcommand{\nPt}{nP_{0.2}}
\newcommand{\Pt}{P_{0.2}}
\newcommand{\kt}{k_{0.2}}

\newcommand{\Tsky}{\bar{T}_{\rm sky}}
\newcommand{\Ta}{\bar{T}_a}
\newcommand{\Tsig}{\bar{T}_{\rm sig}}

\newcommand{\Poh}{P_{\rm HI}}

\newcommand{\Pshot}{P_{\rm shot}}
\newcommand{\Nfeed}{N_{\rm feed}}
\newcommand{\Nyear}{N_{\rm year}}
\newcommand{\Asur}{A_{\rm survey}}
\newcommand{\Lcyl}{L_{\rm cyl}}
\newcommand{\Wcyl}{W_{\rm cyl}}

\newcommand{\Vsur}{V_{\rm survey}}

\newcommand{\ie}{i.e.}
\newcommand{\Nb}{{\it N}-body}
\newcommand{\Nbody}{$N$-body}

\def\prd{Phys. Rev. D}
\def\apj{ApJ}
\def\aap{A \& A}
\def\mnras{MNRAS}
\begin{document}
\setlength{\parskip}{0pt}
\title{The foreground wedge and 21 cm BAO surveys}
\author[Seo \& Hirata]{
Hee-Jong Seo $^{1,2}$ \& Christopher M. Hirata $^{1}$\\
$^{1}$ Center for Cosmology and Astroparticle Physics, Department of Physics, The Ohio State University, OH 43210, USA\\
$^{2}$ Department of Physics and Astronomy, Ohio University, Clippinger Labs, Athens, OH 45701
}

\date{\today} 
\pagerange{\pageref{firstpage}--\pageref{lastpage}} \pubyear{2011}
\maketitle
\label{firstpage}

\begin{abstract}
Redshifted H{\sc\,i} 21 cm emission from unresolved low-redshift large scale structure is a promising window for ground-based Baryon Acoustic Oscillations (BAO) observations.
 A major challenge for this method is separating the cosmic signal from the foregrounds of Galactic and extra-Galactic origins that are stronger by many orders of magnitude than the former. The smooth frequency spectrum expected for the foregrounds would nominally contaminate only very small $\kpar$ modes; however the chromatic response of the telescope antenna pattern at this wavelength to the foreground introduces non-smooth structure, pervasively contaminating the cosmic signal over the physical scales of our interest.
Such contamination defines a wedged volume in Fourier space around the transverse modes that is inaccessible for the cosmic signal.
 In this paper, we test the effect of this contaminated wedge on the future 21 cm BAO surveys using Fisher information matrix calculation. We include the signal improvement due to the BAO reconstruction technique that has been used for galaxy surveys and test the effect of this wedge on the BAO reconstruction as a function of signal to noises and incorporate the results in the Fisher matrix calculation. We find that the wedge effect expected at $z=1-2$ is very detrimental to the angular diameter distances: the errors on angular diameter distances increased by 3-4.4 times, while the errors on $\hz$ increased by a factor of 1.5-1.6.  We conclude that calibration techniques that clean out the foreground ``wedge'' would be extremely valuable for constraining angular diameter distances from intensity-mapping 21 cm surveys. 
 \end{abstract}


\section{Introduction}
\label{sec:intro}

Baryon acoustic oscillations (BAO) imprinted in the large scale clustering of matter have emerged as one of the leading methods to measure the effect of the mysterious dark energy on the geometry of the Universe \citep[e.g.][]{Hu96,Eisen03,Blake03,Linder03,Hu03,SE03}.
BAO were formed by primordial sound waves that had propagated through tightly coupled photons and baryons in very early Universe and subsequently freezed out at the epoch of recombination when the photons and baryons decouped.
The distance that the sound wave has traveled before the recombination epoch is called the sound horizon scale, and this sets the physical size of this feature. The sound horizon scale straightforwardly depends on the sound speed and the time the sound wave traveled, which are precisely calibrated by the cosmic microwave background data. Therefore the observed sizes of BAO in comparison to this true physical size construct a robust standard ruler test, providing cosmological distances, such as angular diameter distances and Hubble parameters to various redshifts, i.e., the expansion history of the Universe.
 
The sound horizon scale today is $\sim 150\Mpc$ and BAO surveys have been designed to measure the large scale structure over many hundreds of $\Mpc$ or a few $\Gpc$ in order to reduce statistical sample variance in the measurements of the $150\Mpc$-long BAO scale. BAO surveys historically have focused on the optical band using spectroscopic or photometric methods, where we measure the large scale clustering in galaxies (or quasars and LyA forests) as biased tracers of the underlying dark matter  \citep[e.g.][]{2005MNRAS.362..505C, 2005ApJ...633..560E, 2007MNRAS.381.1053P, 2010MNRAS.401.2148P, 2011MNRAS.415.2892B, 2011MNRAS.418.1707B, 2011MNRAS.416.3017B, 2012MNRAS.427.2132P, 2012MNRAS.427.2146X, 2012ApJ...761...13S, 2013MNRAS.435...64K, 2014MNRAS.441...24A, 2014MNRAS.441.3524K,2014JCAP...05..027F,2015MNRAS.449..835R,2015A&A...574A..59D}.
 
 Redshifted 21 cm emission from neutral hydrogen has long been investigated as a probe for the epoch of reionization. However, even after reionization neutral gas can still be found in galaxies, and thus intensity mapping in the 21 cm line represents a possible route to low and intermediate-$z$ measurement of the BAO signal \citep[e.g.,][]{2006astro.ph..6104P, 2008PhRvL.100i1303C, 2008arXiv0807.3614A,Wolz14,Bull15}. Among many potential advantages of the 21cm surveys, we note the large instantaneous field of view and hence fast survey speed, comparatively low cost of the telescope, feeds, and electronics, a spectral (hence redshift) resolution that can be very high since it is determined by digital signal processing rather than optics, and absence of the complex atmospheric absorption and emission features that present a challenge in the far-red spectral domain. Such advantages and promising scientific return have motivated ongoing efforts for near future 21cm BAO experiments such as the Canadian Hydrogen Intensity Mapping Experiment (CHIME)~\footnote{http://www.mcgillcosmology.ca/chime}, the Baryon Acoustic Oscillation Broadband and Broad-beam~\citep[BAOBAB,][]{2013AJ....145...65P}, BAORadio~\citep{2012CRPhy..13...46A}, the BAO from Integrated Neutral Gas Observations experiment~\citep[BINGO,][]{2013MNRAS.434.1239B}, Ooty~\citep{2014JApA...35..157A} and others. 
 
The biggest challenge of this method is separating the astrophysical foreground emissions from the cosmic signal that is weak by many orders of magnitudes than the former. The good news is that the foreground due to Galactic and extragalactic synchrotron and free-free emission are spectrally smooth, which means they in principle would only contaminate very small line-of-sight wave modes in Fourier space (i.e., mainly fluctuations in the transverse direction). Losing the small line-of-sight Fourier modes \citep[i.e.,  $\kpar<0.01-0.02\hMpc$, references in ][]{2008PhRvL.100i1303C} has very little impact on the BAO error forecasts and therefore many, if not all, error forecasts for the 21cm BAO surveys assume a perfect foreground subtraction taking this advantage and include the foreground effect mainly in the error budget \citep[e.g.,][]{2010ApJ...721..164S,Bull15}.

In reality, however, the frequency dependence
of an interferometer's fringe pattern (i.e., chromatic instrument
response) can cause the spectrally smooth foreground
spectrum to ring at a different angular location as a function of frequency. This introduces an un-smooth foreground component and therefore makes a clean separation between
the foreground and the cosmic signal very difficult \citep[e.g.,][]{Alonso15}.
However, it does turn out to be confined to a so-called ``foreground wedge'' at low $\kpar$ and high $\kperp$ \citep{2010ApJ...724..526D, 2012ApJ...745..176V, 2012ApJ...752..137M}, leaving a cleaner ``window'' at high $\kpar$ and low $\kperp$. 
The wedge effect is expected to be much alleviated for surveys with very small side lobes, such as single dish surveys with low aberrations and unobstructed apertures (e.g. the surveys performed at the Green Bank Telescope \citep{Masui13}). For interferometer arrays, side lobes due to various instrumental effects that scale with frequency will introduce the foreground wedge.
The wedge is not a fundamental effect, in the sense that with excellent baseline-to-baseline calibration the foregrounds in this region can be removed, but the factor by which {\em raw} systematics dominate over cosmological signal is orders of magnitude greater in the wedge than the window (see e.g. \citealt{2015arXiv150207596T} for a discussion in the context of epoch of reionization studies). It is thus of great interest to understand the impact of the wedge on cosmological constraints. {\em How are cosmological constraints affected if we assume that only the cosmological signal in the foreground ``window'' is available?} Or, stated more optimistically, {\em what are the potential gains to cosmology if advances in side-lobe mapping and data processing techniques can clean out the wedge?} \citet{2015MNRAS.447.1705P}, for example, has shown the impact of the foreground wedge on redshift-space distortion measurements, which is destructive at $z\sim 8$ but can place non-trivial constraints at $z\sim 1$. 

The purpose of this paper is to estimate the effect of removing the wedge on the BAO analysis. We use the Fisher information matrix calculation as well as \Nb\ simulations to consider the wedge effect on the BAO reconstruction.

The expected performance of a 21 cm BAO survey in the presence of the wedge effect could be straightforwardly calculated using the Fisher matrix formalism by removing the contribution from the Fourier modes in the wedged volume. Such calculation has a very simple dependence on the expected sample variance and the expected strength of the BAO signal from the assumed large-scale clustering data. The strength of the BAO signal is reduced at low redshift due to various nonlinearities in the process of structure growth and redshift-space distortions in a predictable manner. Fortunately, a large portion of such nonlinear reduction of the signal can be undone by using the BAO reconstruction technique \citep{2007ApJ...664..675E}. In this technique one uses the observed clustering of a given tracer to estimate the gravitational potential field due to the underlying matter density field and use the estimated potential field to correct for the motions of mass elements that weakened the BAO signal. The BAO reconstruction is proven to be successful in subsequent simulations \citep[e.g.,][]{Huff07, 2008ApJ...686...13S,2009PhRvD..79f3523P,2009PhRvD..80l3501N,2010ApJ...720.1650S,2011ApJ...734...94M} as well as in recent galaxy redshift surveys \citep[e.g.,][]{2012MNRAS.427.2132P,2012MNRAS.427.3435A}, and now has become a standard tool for BAO surveys with a good redshift precision. 

In order to include the effect of the BAO reconstruction in the Fisher matrix formalism, we need an estimate of the strength of the BAO signal after reconstruction. In a large signal-to-noise limit (per Fourier mode or per pixel), we conventionally assume that the width of the BAO peak reduces to about a half after reconstruction at  $z\sim 0.3$ \citep{2007ApJ...664..675E,2007ApJ...665...14S}. 21cm BAO surveys, however, operate often at a low S/N per mode due to the high system (foreground as well as instrument) temperature relative to the weak line signal, while the total S/N contributed by many modes is boosted by the large survey area. The reconstruction efficiency is in general expected to decrease as the noise level of the density field increases; \citet{2008ApJ...686...13S} and \citet{2011ApJ...734...94M} used simulations to show that a density field with $S/N$ per mode (at $k=0.2\ihMpc$) as low as $\sim 0.6$ nevertheless shows a non-negligible improvement by the reconstruction; recently, \citet{2014MNRAS.445.3152B} tested a regime with $S/N \gtrsim 0.2$ for galaxy surveys and found a similar result. In the current paper, we test a density field with $S/N$ as low as 0.1.

An analytic prediction on the BAO reconstruction performance as a function of (shot) noise level has been derived based on the Lagrangian Perturbation Theory formalism by \citet{2010arXiv1004.0250W} in real space. The presence of the wedge contamination makes estimating the effect of the BAO reconstruction using such an analytic approach nontrivial especially when including redshift-space distortions. In this paper, we investigate the effect of the wedge on the BAO reconstruction using \Nb\ simulations as a function of a signal-to-noise ratio per mode as low as $\sim 0.1$. We incorporate our results in the Fisher matrix calculation and derive the final effect of the wedge contamination on the distance precisions expected for 21 cm BAO surveys. 

This papers is structured as follows. In \S~\ref{sec:Fisher}, we set up Fisher matrix calculations in the presence of the wedge effect as well as introducing a fiducial 21cm BAO survey that we adopt from \citet{2010ApJ...721..164S}. In \S~\ref{sec:BAOrec}, we present the details of performing BAO reconstruction in the presence of the wedge at different redshifts and at various signal to noise levels that are simulated. In \S~\ref{sec:Results}, we present the results of this paper, the BAO reconstruction efficiency at different signal-to-noise levels as well as at different extents of the wedge, and the Fisher matrix calculations incorporating BAO reconstruction. Finally, in \S~\ref{sec:Con}, we conclude.

\section{Setting up the Fisher matrix analysis for 21cm BAO surveys}
\label{sec:Fisher}

The Fisher matrix formalism has been an effective tool for predicting constraints on cosmological parameters for future BAO surveys.
In this section, we lay out the components of the Fisher matrix calculations aimed at incorporating the wedge effect. We start from the method in \citet{2007ApJ...665...14S} that isolates the BAO-only information. In order to quantify the effect of the wedge, we adopt the same fiducial 21cm survey configuration described in \citet{2010ApJ...721..164S} and compare the results before and after incorporating the wedge effect. 

\subsection{Fiducial cosmology of the Fisher analysis}
For simplicity, we define the fiducial cosmological parameters to be $\Om = 0.27$, $h=0.72$, $\OL=0.73$, $\Ok=0$, $\Ob = 0.0238$, $\tau = 0.17$, $n_s = 0.99$, which is one of the cosmologies used\footnote{This was based on the first year Wilkinson Microwave Anisotropy Probe \citep{2003ApJS..148..175S}} in \citet{2007ApJ...665...14S}, rather than using the best fit cosmologies from the most up-to-date CMB probes. We are therefore testing the effect of the wedge given this cosmological assumption; however, we do not believe that the {\it relative} effect of the wedge will sensitively depend on this assumption. For example, the wedge angle is not much different between the fiducial cosmology and the Planck ($TT$ spectra + polarization at low multipoles) best fit  $\LCDM$  cosmology \citep{2015arXiv150201589P}; the cosine of the angle is different only by $~3\%$ (Figure 1).

\subsection{Fiducial 21cm survey configuration}

When investigating the wedge effect, we assume a fiducial 21cm survey in order to approximately simulate a reasonable redshift range as well as the signal-to-noise range to consider. We adopt the fiducial CRT survey assumed in \citet{2010ApJ...721..164S} that represented one possible example of cost-effective future 21cm BAO survey designs. Table \ref{tab:fid} in this paper is taken from Table 1 in \citet{2010ApJ...721..164S} that defines the fiducial survey configuration. While we refer interested readers to \citet{2010ApJ...721..164S}  for more details on assumptions, we summarize the key aspects here.

\begin{table}
\caption{Fiducial CRT configuration from Table 1 of \citet{2010ApJ...721..164S}. For technical reasons, it was assumed that the redshift range is covered by two distinct configurations of the telescope. The fiducial telescope is a square in the sense that the total width $N_{\rm cyl}\Wcyl$ is equal to $\Lcyl$.}
\centering
\begin{tabular}{lcc}
\hline
\hline
Parameter &  Low redshift  & High redshift \\ \hline
Redshift range & $0.66-1.24$  & $1.22-2.11$ \\  
Number of cylinders, $N_{\rm cyl}$ & 7 & 10 \\
Length of Cylinder, $\Lcyl$ (m)&	99.8	&	142.8 \\
Feed spacing, $d_F$ (m)		&	0.39	&	0.558 \\
Width of Cylinder, $\Wcyl$ (m)  & 	14.3	&	14.3  \\	 
Duty factor, $D_f$    	&	0.5	&	0.5	\\
$\Nyear$ (years)	&	1.40	&	0.87	\\
$x_{\rm HI}\Omega_{\rm H,0}$			& 	0.00037	&	0.00037	\\
bias			&	1.0	&	1.0 \\
Sky temperature, $\Tsky$ (K)		& 	10	&	10 \\
Antenna temperature, $\Ta$  (K)		& 	50	&	50	\\
gain, g			& 	0.8	& 0.8	\\
$\Pshot$		&	100.0	& 100.0	\\
\hline
\label{tab:fid}
\end{tabular}
\end{table}

The signal to noise of the BAO feature in the 21cm intensity power spectrum when averaged over $dk$ and $d\mu$, where $k$ is a wavenumber and $\mu$ is the cosine of the line-of-sight angle, depends on the strength of the 21cm signal due to the large scale clustering of the neutral hydrogen as signal and instrumental noise and the sky background \footnote{We include here the BAO damping factor that is not explicit in the equation in \citet{2010ApJ...721..164S}.}\footnote{we assume a Gaussian density field; Takahashi et al. (2011) and Ngan et al. (2012) have shown that the non-Gaussian errors on the BAO measurement in multi-parameter fitting are negligible.}:
\begin{equation}
\frac SN = \sqrt{\frac{2\pi k^2\,dk\,
    d\mu\,\Vsur}{2(2\pi)^3}}
    \frac{\Poh(k,\mu)\,e^{-(\kperp^2\Sigxy^2+\kpar^2\Sigz^2)/2}\,\hat W^2}{\Poh(k,\mu)\hat{W}^2 + P_{\rm th}(k,\mu) + P_{\rm shot}},
\label{eq:StoN}
\end{equation}
where $\Vsur$ is the total volume of the survey and we add a galactic shot noise contribution $P_{\rm shot} = 1/\bar n$ due to the discreteness effect of the H{\sc\,i} sources with an effective number density $\bar{n}$ \citep{2010ApJ...721..164S}. $\Sigxy$ and $\Sigz$ are the Gaussian damping lengths that account for the degradation of the BAO signal strength due to various nonlinearities. Due to redshift-space distortions along the line of sight, $\Sigz$ tends to be larger than $\Sigxy$. The values of $\Sigxy$ and $\Sigz$ can be fairly precisely predicted as a function of cosmology and redshift in linear theory \citep[e.g.,][]{2007ApJ...664..660E,Mat08}.

For a dual-polarization cylinder telescope built near the Equator and with the long axis of the cylinders oriented North-South, the instrument thermal noise term, expressed in conventional galaxy power spectrum units ($h^{-3}$ Mpc$^3$), is
\begin{equation}
P_{\rm th}(k,\mu) = \left( \frac{g\Tsky+\Ta}{g\Tsig }\right)^2
\frac{dV}{dA\,df} \frac{2\pi\lambda_{\rm obs}\Wcyl}{\Lcyl^2 D_f \Nyear}, 
\end{equation}
where $\Tsig$ is the average brightness temperature due to the 21cm line, $\Tsky$ is the average sky temperature (e.g., due to foregrounds), $\Ta$ is the average antenna noise temperature or the amplifier noise temperature, and $\lambda_{\rm obs} = c(1+z)/f_{\rm 21cm}$ is the observed wavelength \citep{2010ApJ...721..164S}. The conversion factor $dV/dA\,df$ is from observed angle-frequency volume element to comoving cosmic volume, and has units of $(h^{-1}\,{\rm Mpc})^3$/sr/Hz. This accounts for the distribution of time over the full range of right ascension ($2\pi$) and the declination range given by the Nyquist limit of the beam, $|\sin{\rm Dec}|<\lambda_{\rm obs}/(2d_F)$, where $d_F$ is the feed spacing; the survey area is
\begin{equation}
\Asur = 2\pi\frac{\lambda_{\rm obs}}{d_F}=2\pi{\lambda_{\rm obs}}\frac{\Nfeed}{\Lcyl},\label{eq:Area}
\end{equation}
so long as $d_F>\lambda_{\rm obs}/2$ (as is the case here -- otherwise the telescope is sensitive all the way to the celestial poles). The number of feeds is $\Nfeed = \Lcyl/d_F$. We have incorporated a duty factor $D_f\le 1$ (the fraction of time the telescope is operating in science mode) and a survey duration $\Nyear$.

We assume that the neutral hydrogen traces the matter, so that $\Poh(k,\mu)=b^2(1+\beta\mu^2)^2P_m(k)$ with bias $b=1$ and redshift-space distortion parameter \citep{1987MNRAS.227....1K} $\beta = d\ln G(a)/d\ln a/b$ where $G(a)$ is the growth function.  Note that assuming $b>1$ improves the signal to noise. The signal temperature is \citep{2007RPPh...70..627B,2008PhRvD..78j3511P,2008PhRvL.100i1303C}
\begin{eqnarray}
\Tsig=188 \frac{x_{\rm HI}(z)\Omega_{\rm H,0} h (1+z)^2}{H(z)/H_0} {\rm\,mK},
\end{eqnarray}
where $x_{\rm HI}$ is the neutral hydrogen fraction at $z$, $\Omega_{\rm H,0}$ is the ratio of the hydrogen mass density to the critical density at $z=0$, and $H(z)$ is the Hubble parameter at $z$.

The signal is suppressed by an instrument response function\footnote{This response function is often included in the noise rather than the signal, in which case the noise power is inversely proportional to the number of baselines at the relevant separation. Where one should put it depends on whether the sky map under consideration is instrument-convolved or not; the choice has no impact on the signal-to-noise ratio or cosmological parameter constraints, so long as it appears somewhere!}, which will be modified in the next section to include the wedge effect. The general response function is
\begin{equation}
\hat W({\bmath k}) = A(u\lambda_{\rm obs}, v\lambda_{\rm obs})
\left[\frac{\sin(\tau\Delta f/2)}{\tau\Delta f/2}\right]^2,
\end{equation}
where $(u,v,\tau)$ is the Fourier mode in observed (angle-frequency) space that corresponds to ${\bmath k}$ (the conversions are given explicitly in \S\ref{ss:wedge}), and $\Delta f$ is the width of a frequency channel (i.e. the pixel transfer function in the radial direction). The first term is the distribution of baselines at separation $(u\lambda_{\rm obs}, v\lambda_{\rm obs})$, normalized to $A(0,0)=1$. We assume that the frequency channels are narrow ($\ll 1$ MHz) so that the last term can be neglected.

This response function is slightly different from the form adopted in \citet{2010ApJ...721..164S}. \citet{2010ApJ...721..164S} assumed a specific choice of pixelization along the declination direction and accounted for the effect of the default number of cylinders that fill the baseline along the right ascension; here we assume a generalized response function approximation in both directions for simplicity.  This gives 
 \begin{equation}
\hat{W}_{\rm inst}({\bmath k}) = (1-| k_{\rm x}/\kmax| )  (1-|k_{\rm y}/\kmax|) \label{eq:window}
\end{equation}
for $k_{\rm x}, k_{\rm y} \leq \kmax$, and otherwise 0. Here $k_{\rm x}$ is a wavevector component in the direction of right ascension, $k_{\rm y}$ is a component in the direction of declination, and $\kmax$ is set to be $2\pi \Lcyl/\lambda_{\rm obs}/D(z)$, where $D(z)$ is the comoving angular diameter distance. For example, $\kmax = 0.35\ihMpc$ at $z=1$ and $0.6\ihMpc$ at $z=2$ at our fiducial cosmology. Although not presented in this paper, we repeated the entire analysis in this paper with the response function presented in \citet{2010ApJ...721..164S}; qualitatively we find very similar results, while quantitatively the current response function produces larger errors.\footnote{\citet{2010ApJ...721..164S} adopted  \\$\hat{W}_{\rm inst}({\bmath k}) = \exp \left[-1.5(k_{\rm x}/\kmax')^2\right]\Theta(\kmax'-k_{\rm y})$ \\with $\kmax'=\pi \Lcyl/\lambda_{\rm obs}/D(z)$. }
As a caveat, although we do not put a limit on $\kmax$, all our Fisher matrix calculations do not account for a BAO information beyond $0.5\ihMpc$.

\subsection{The wedge effect due to the foreground contamination}
\label{ss:wedge}

The wedge effect becomes important when we lift the assumption that the side lobes are negligible. We first discuss some considerations of the foreground before the assumption is lifted. 

The foreground subtraction for all cosmological 21 cm experiments -- whether low-$z$ for BAO or high-$z$ for the epoch of reionization -- depends on the spectral smoothness of synchrotron and free-free
emission \citep{2004MNRAS.355.1053D, 2004ApJ...608..611G, 2004ApJ...615....7M, 2004ApJ...608..622Z, 2005ApJ...625..575S, 2006ApJ...653..815M}, as compared to the 
cosmological signal for which the large-scale clustering is expected to
introduce variations in the H{\sc\,i} density and temperature as a function
of observed frequency (or distance from the observer).
Foreground mechanisms with strong frequency-dependent structure, such as extragalactic radio recombination lines, are expected to be small \citep[e.g.][]{2003MNRAS.346..871O}. This phenomenon can be quantified by considering the cosmological data cube in comoving coordinates $(r_x,r_y,r_z)$ (with $r_z$ being the line-of-sight or redshift direction) and mapping it to the directly observable quantities $(\theta_x,\theta_y,f)$ (with $\theta$ being the angular direction on the sky in orthographic projection with the primary beam line of sight at $\theta_x=\theta_y=0$, and $f$ as the observed frequency: $f = f_{\rm 21 cm}/(1+z)$, where $f_{\rm 21cm}=1.42$ GHz is the rest-frame frequency of the 21 cm line). Therefore there is a Jacobian mapping between these spaces, which is locally diagonal:
\begin{equation}
\frac{\partial r_x}{\partial\theta_x} = \frac{\partial r_y}{\partial\theta_y} = D(z)
\label{eq:Jac1}
\end{equation}
and
\begin{equation}
\frac{\partial r_z}{\partial f} = \frac{ -c(1+z)^2}{f_{\rm 21cm} H(z)} = \frac{-(1+z)\lambda_{\rm obs}}{H(z)},
\label{eq:Jac2}
\end{equation}
where $D(z)$ is the comoving angular diameter distance, and $H(z)$ is the Hubble rate. A foreground signal that exists in $(\theta_x,\theta_y,f)$-space can be characterized by considering its power spectrum, which lives in the Fourier-conjugate space $(u,v,\tau)$, i.e. the $(u,v,\tau)$ Fourier mode contributes an intensity $I(\theta_x,\theta_y,f)\propto e^{2\pi i(u\theta_x+v\theta_y + \tau\nu)}$. As a caveat, $\tau$ is not exactly a local Fourier conjugate of $f$ (often referred to as $\eta$) in a wide-field instrument except at the center of the field because $u$ and $v$ are dependent on $f$ except at the center. Given the mapping of Eqs.~(\ref{eq:Jac1},\ref{eq:Jac2}), a Fourier mode in the observable cube maps into a Fourier mode in the cosmological cube according to
\begin{equation}
k_x = \frac{2\pi u}{D(z)}, ~~~~
k_y = \frac{2\pi v}{D(z)}, ~~~~{\rm and}~~~~
k_z = \frac{-2\pi \tau H(z)}{(1+z)\lambda_{\rm obs}}.
\label{eq:cosmo-cube}
\end{equation}
In the context of a radio interferometer, the angular conjugate space $(u,v)$ corresponds to transverse baseline separation in units of $\lambda_{\rm obs}$, and $\tau$ to time delay. (The identification of conjugate-frequency with time delay is a consequence of the fact that the power spectrum of the electric field from a source and its correlation function are Fourier transforms; see e.g. \citealt{2012ApJ...756..165P} for a rigorous treatment.) A foreground that is spectrally smooth in $f$ contaminates only the small (in absolute value) conjugate-frequencies $\tau$, and hence small $k_z=\kpar$ in the cosmological box.

Since there are not many Fourier modes available for small $\kpar$, removing the small-$\kpar$ modes has only a small impact on the predicted precisions. But while the main astrophysical foregrounds are indeed spectrally smooth, if an antenna has a far side-lobe response then the cross-correlation of the electric fields measures with two such antennas can have a contribution at long time delay $\tau$ (or in frequency space, it can vary rapidly as a function of $f$). We summarize the derivation of the effect here, drawing extensively on derivations in the literature \citep{2010ApJ...724..526D, 2012ApJ...745..176V, 2012ApJ...752..137M, 2012ApJ...756..165P, 2012ApJ...757..101T, 2014PhRvD..90b3018L, 2014PhRvD..90b3019L}.

For two antennas separated by distance $L$ (assumed perpendicular to the target field, so that $\sqrt{u^2+v^2} = L/\lambda_{\rm obs}$), the cosmological transverse wave number being sampled is
\begin{equation}
k_\perp = \sqrt{k_x^2+k_y^2} = \frac{2\pi\sqrt{u^2+v^2}}{D(z)} = \frac{2\pi L}{\lambda_{\rm obs}D(z)}.
\end{equation}
A foreground in a far side-lobe produces a time-delayed signal in the two antennas according to the geometric propagation time, which may be as large as $L/c$, so a contaminated region in the Fourier domain is produced at $|\tau|\le L/c$. Mapped from the observable cube into the cosmological cube, this corresponds to
\begin{equation}
|\kpar|=|k_z|= \frac{2\pi|\tau|H(z)}{(1+z)\lambda_{\rm obs}}
\le \frac{2\pi L H(z)}{c(1+z)\lambda_{\rm obs}}
= \frac{D(z)H(z)}{c(1+z)}k_\perp.
\label{eq:Wedge}
\end{equation}

Equation~(\ref{eq:Wedge}) defines a ``wedge'' in Fourier space that is contaminated by the response to sources far from the main beam due to geometric time delay, and a conical\footnote{In 3D Fourier space $(k_x,k_y,k_z)$, Eq.~(\ref{eq:Wedge}) defines two cones, one at $\kpar>0$ and one at $\kpar<0$. Since power spectra satisfy the mathematical identity $P({\bmath k})=P(-{\bmath k})$, one normally considers only positive $\kpar$.} ``window'' in which this process produces no contamination.

Some words of warning are in order before we proceed. The wedge does {\em not} represent information that is fundamentally irretrievable, since with excellent knowledge of the side-lobe response and maps of the radio sky one could compute the contribution from sources far from the main beam and subtract it off (or solve for the contaminating sources as part of a global model). The ``windows'' are also not perfectly clean: other effects, notably reflections in cables or antennas, can lead to a sky signal received at time $t$ being detected again at some later time $t+\Delta t$, and thus producing spurious power at $\tau=\Delta t$. A similar effect can be produced by linear polarization-to-intensity (Stokes $Q,U\rightarrow I$) leakage in the presence of Galactic Faraday rotation: the arrival times of the left- and right-circularly polarized pulse from a distant electron are separated by a relative delay $\Delta t_{\rm L-R}$ because of the different indices of refraction for the two circular polarizations, which produces contamination in the Fourier modes corresponding to $\eta = \Delta t_{\rm L-R}$ \citep[e.g.,][]{2003A&A...404..233H,2003A&A...403.1031H,2006AN....327..487D,2009MNRAS.399..181P, 2009A&A...500..965B,2010A&A...522A..67B, 2013ApJ...769..154M} \citep[c.f., for a removal technique, ][]{2014ApJ...781...57S,2015PhRvD..91h3514S}. 

However, multiple recent 21 cm experiments, including the PAPER \citep{2013ApJ...768L..36P} and MWA \citep{2014PhRvD..89b3002D, 2015arXiv150207596T} instruments that have produced the tightest upper limit on the reionization signal, have found that the wedge effect is the brightest foreground-instrument interaction signal in their data, and is orders of magnitude brighter than thermal noise. Therefore in this paper we forecast the effect of the wedge on BAO measurements using the 21 cm intensity mapping technique. The basic question is to determine how much loss of precision occurs if the wedge is removed, or alternatively what gains would be realized with calibration techniques that clean out the wedge. In what follows, we address this question by considering BAO forecasts with two different cases -- a ``Default'' case, assuming the full 3D Fourier domain is available and noise-limited, and a ``Wedge'' case, in which information in the wedge is assumed to be inaccessible.

\subsection{Incorporating the wedge effect into the Fisher matrix: case of no reconstruction}

\begin{figure}
  \centering
  \includegraphics[width=0.8\linewidth]{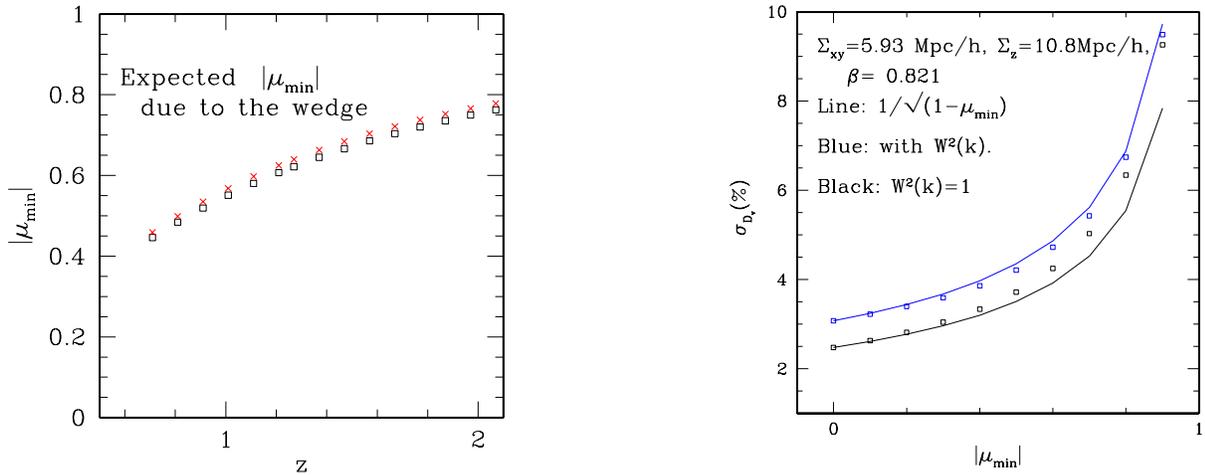}
\caption{The minimum cosine line-of-sight angle $\mumin$ set by the wedge effect as a function of redshift given our fiducial cosmology (black squares); red crosses show the minimum angle assuming the Planck $\LCDM$ best fit cosmology \citep{2015arXiv150201589P}. Note that wedge effect solely depends on the cosmological parameters rather than the telescope configuration. Assuming WMAP1, $\mumin=0.55$ at $z=1$ and 0.76 at $z=2$.}\label{fig:zmumin}
\end{figure}

\begin{figure}
\centering
\includegraphics[width=0.8\linewidth]{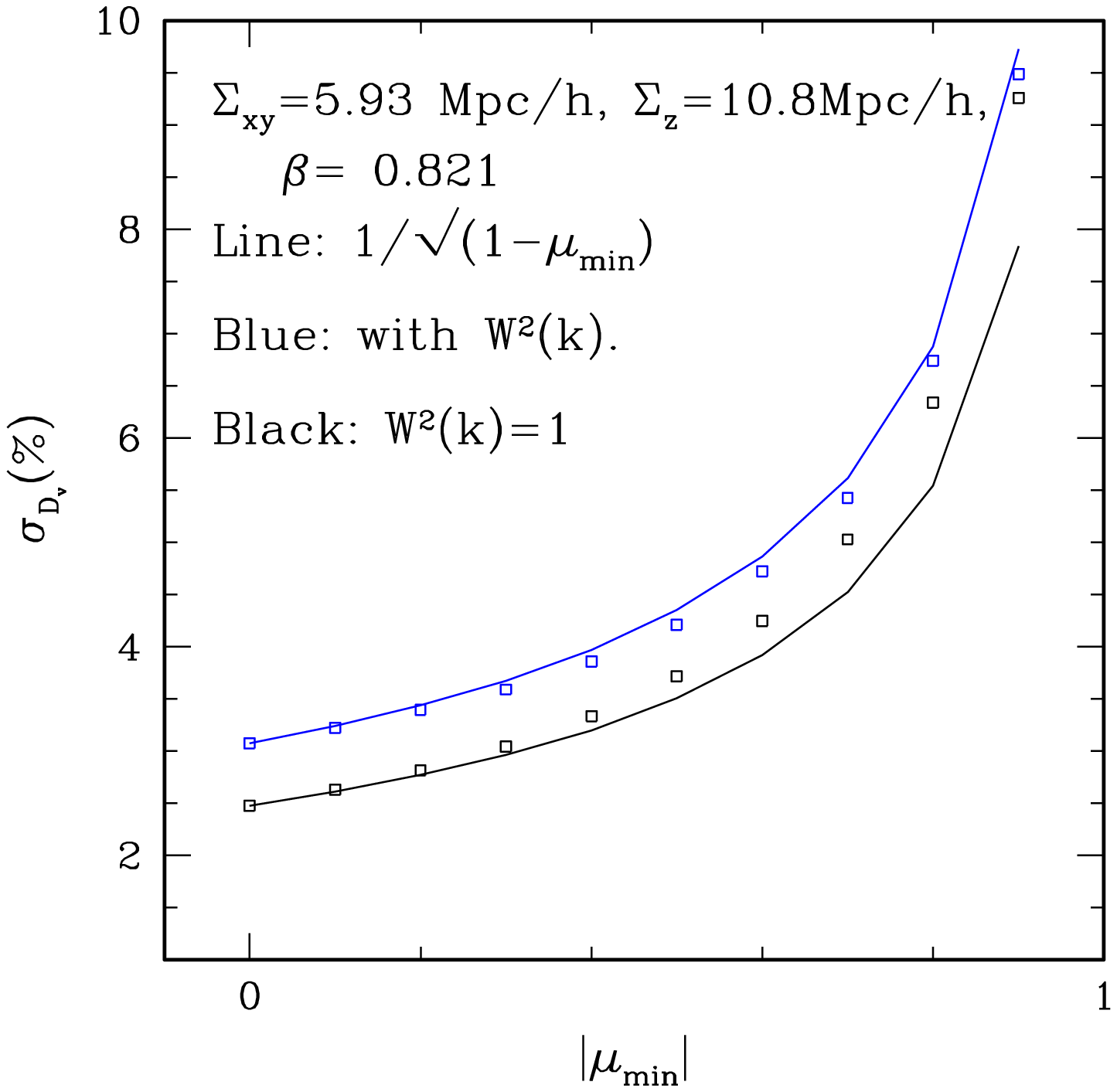}
\caption{The effect of $\mumin$ on an isotropic distance scale $\DV$ (squares)
  in comparison to what is expected based on the number of Fourier
  modes cut off as a function of $\mumin$ (lines). We assume a
  condition similar to our fiducial
  redshift bin at $z=1$ except for the volume, i.e., we assume $\Sigxy = 5.93$, $\Sigz=10.8$,
$\beta=0.82$, $\kmax=0.6\ihMpc$ (in Eq. \ref{eq:window}), and $\nPt =
0.46$ while assuming a volume of $1\trihGpc$. The reference line is the
expected functional dependence of $\sDV$ based on the number of
Fourier modes cut off by $\mumin$. Blue points/line include the triangular window function effect, while the overlaid black points/line assume no window function.  }\label{fig:Dvmumin}
\end{figure}

It is straightforward to include the wedge in the BAO Fisher matrix in the absence of reconstruction.
We simply incorporate the effect of the wedge by excluding all the signal-to-noise
contributions (Eq. \ref{eq:StoN})
from the  Fourier modes at $\mu < \mumin$ when calculating the Fisher matrix:
\begin{equation}
\mumin = \frac{\kpar}{\sqrt{\kperp^2+\kpar^2}}=\frac{D(z)H(z)/[c(1+z)]}{\sqrt{1+\{D(z)H(z)/[c(1+z)]\}^2}}.
\label{eq:mumin}
\end{equation}
Equivalently, we may replace the response (i.e., window) function in Eq.~(\ref{eq:window}) with
\begin{eqnarray}
\hat{W}({\bmath k}) = \hat W_{\rm inst}({\bmath k})\Theta({\mu-\mumin}),
\label{eq:windowf}
\end{eqnarray}
where $\mumin$ is the cosine of the wedge angle from Eq.~(\ref{eq:mumin}). In the absence of reconstruction, the two formalisms are equivalent, but when we study reconstruction, the window function formalism will be more useful due to its flexibility in possibly softening the wedge after reconstruction. For simplicity, we  assume the same $\mumin$ cutoff before and after reconstruction. 
Again, note that the minimum cosine angle $\mumin$ is a purely function of cosmology and the redshift of the aimed signal such as $H(z)$ and $D(z)$, not depending on the configurations of the telescope at all. Using our fiducial cosmology, $\mumin =0.55$ at  $z \sim 1$ and 0.76 at $z \sim 2$.  
Figure~\ref{fig:zmumin} shows the $\mumin$ as a function of
redshift over our fiducial range $0.7\lesssim z \lesssim 2$.
Figure~\ref{fig:Dvmumin} shows the effect of $\mumin$ on the precision
of the isotropic distance scale $\DV$
\footnote{$\DV$ is calculated by deriving Fisher matrix of $D(z)$ and $H(z)$  and then contracting the Fisher matrix of $D(z)$ and $H(z)$ with the vector $(1, -1)$ assuming radial and transverse distances scale the same way, as explained in \citet{2007ApJ...665...14S}. This $\DV$ constraint is therefore equivalent to varying $D_V$ with fixed Alcock-Pazynski \citep{1979Natur.281..358A} shape $D_AH$, corresponding to a BAO distance scale derived from a spherically averaged clustering information.} 
as a sanity check assuming our fiducial
redshift bin at $z=1$; $\sDV$ as a function of 
$\mumin$ is roughly consistent with the expected dependence based on the
number of Fourier modes that are lost, i.e., $\propto 1/
\sqrt{1-\mumin}$; the deviation at high $\mumin$ is probably because
the effective damping length also increases as $\Sigz$ becomes more important than $\Sigxy$ with increasing
$\mu$. Overall the wedge effect are consistent with the mode counting when other conditions of the survey is fixed. The blue points/line account for the triangular window function we defined, and the black points/line do not include any window function. The results using the window function from \citet{2010ApJ...721..164S} give a prediction in between the two cases.

\begin{figure*}[b]
  \centering
  \includegraphics[width=0.8\linewidth]{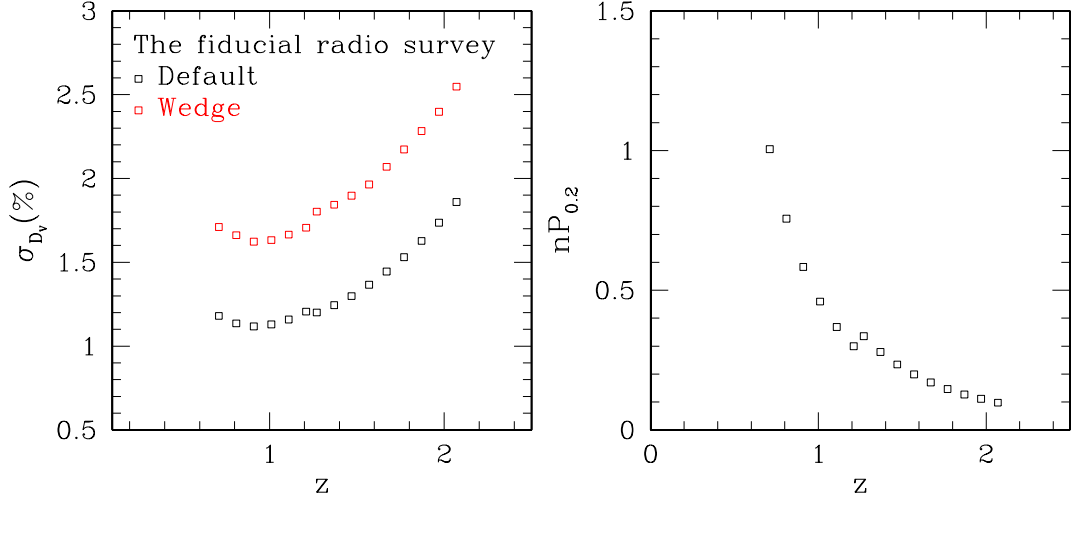}
\caption{Left: error on $\DV$ without (black) and with (red) the wedge effect for the fiducial survey. No BAO reconstruction is assumed here. Therefore, the black squares represent the results similar to those in \citet{2010ApJ...721..164S}. Right: $\nPt$ of the fiducial survey.}\label{fig:sigDvratio}
\end{figure*}

We include $\mumin(z)$ from Figure 1 in the Fisher forecast for our fiducial 21cm BAO survey and investigate the combined effect of the wedge for the survey.  In Figure \ref{fig:sigDvratio}, the
black and red squares show the expected precisions on $\DV$ (i.e.,
$\sDV$) without and with the wedge effect, respectively, based on the
survey configuration in Table \ref{tab:fid}. 

When
the wedge effect is included, we find that $\sDV$ increases about $\sim 36-46$ per cent over the fiducial redshift range. It is interesting that
this ratio is almost constant over the entire redshift range even though the range of $\mumin$ varies
from 0.446 (at $z=0.7$) to 0.762 (at $z=2.0$), which alone would imply a
substantially more increase in $\sDV$ at $z=2$ (i.e., based on $\sqrt{1-\mumin}$, 
 $\sDV$ is expected to increase 30 per cent more at $z=2$ compared to at $z=0.7$). 
Moreover, $\Sigz/\Sigxy$ is proportional to the redshift-space distortions \citep[e.g.,][]{2007ApJ...664..660E} which is larger at high $z$; a larger $\mumin$ at high $z$ would have to use a more damped BAO feature, thereby increasing $\sDV$. 

 The main leverage comes
from the redshift-space distortions that
amplify the power of the Fourier modes along the line of sight more at high $z$; for example, $\beta$ at $z=2$ is $0.94$ while
$\beta$ at $z=0.7$ is $\sim 0.47$ assuming the clustering bias of HI is 1. The discontinuity at
$z\sim 1.22$ in Figure 3 with the wedge effect (red squares) is because the fiducial telescope configuration changes
near this redshift (see Table \ref{tab:fid}) in a way that affects
the modes along the line of sight more. 

The wedge effect we have investigated so far is a combination of Fourier mode counting over the redshift bins of a reasonable fiducial survey without assuming the BAO reconstruction. We next include the BAO reconstruction in the presence of the wedge effect through the BAO damping scales.

\section{The wedge effect on the BAO reconstruction}
\label{sec:BAOrec}

\subsection{BAO reconstruction in Fisher matrix calculations}

The BAO reconstruction technique was first introduced in \citet{2007ApJ...664..675E}
and has been shown to be successful in simulations as well as in
observations
\citep[e.g.,][]{2008ApJ...686...13S,2010ApJ...720.1650S,2009PhRvD..79f3523P,
2009PhRvD..80l3501N, 2011ApJ...734...94M,2012MNRAS.427.2132P,2012MNRAS.427.3435A,2015MNRAS.450.3822W}. The method
assumes a linear continuity equation to estimate the large-scale
displacement field from the gravitational field constructed from the observed
nonlinear density fields of mass tracers (Eq. ~\ref{eq:qvec}). In Fourier space, the continuity equation becomes
\begin{eqnarray}
\hat{\bmath q}({\bmath k})  &=& -\frac{i{\bmath k}}{k^2}S({\bmath k})\hat{\delta}({\bmath{k}}),
\label{eq:qvec}
\end{eqnarray} 
where $\hat{\delta}({\bmath {k}})$ is the Fourier coefficient of the nonlinear 
tracer density field and $\hat{\bmath q}({\bmath{k}})$ is the Fourier
coefficient of the 3-dimensional displacement field. $S(\bmath{k})$ is a smoothing kernel that we convolve with the observed density field to reduce the small-scale nonlinearities before imposing the continuity equation. 

Displacing
the tracers back to the estimated original positions using the derived
displacement field, which are then followed by additional simple
steps, substantially removes the nonlinear mode-coupling effect on BAO
(especially, mode-coupling between large and small wave modes) and reduces the damping of the BAO. We refer interested readers to \citet{2007ApJ...664..675E} and \citet{2009PhRvD..79f3523P} for more details.

The Fisher matrix calculation can include the improvement due to the BAO
reconstruction technique through the nonlinear damping scale,
$\Sigxy$ and $\Sigz$, in Eq.~(\ref{eq:StoN}). Depending on the details of the reconstruction, the level of power spectrum amplitude 
relative to the shot noise may change, but we ignore such effect in this study.
We follow the exactly the same procedure that is summarized in \citet{2011ApJ...734...94M}, 
which was designed to recover the linear redshift-space distortions after reconstruction rather than removing it. 

In galaxy surveys, the conventional assumption is that the BAO
reconstruction halves $\Sigxy$ and $\Sigz$ \citep[e.g.,][]{2008ApJ...686...13S} 
when the signal per Fourier mode is much larger than the
shot noise, \ie, $nP\gg1$ at $k \sim 0.2\ihMpc$ which is the scale relevant to the BAO feature.
We define $\nPt$ in an intensity-mapping radio survey to be the ratio of HI clustering power to the sum of both shot and instrument-thermal noise at this scale:
\begin{equation}
\nPt = \frac{\Poh(k=0.2\ihMpc,\mu=0)}{P_{\rm th}(k=0.2\ihMpc,\mu=0)+\Pshot}.
\label{eq:pnt}
\end{equation}
A low signal-to-noise per mode $\nPt\ll 1$ means that the density field is noisy and therefore the displacement field derived from Eq.~(\ref{eq:qvec}) would be noisy as well, thereby decreasing the efficiency of the BAO reconstruction. Note that we continue to use this definition in the presence of the wedge, even though the $\mu=0$ modes are in the wedge and hence excised.

The right panel of Figure \ref{fig:sigDvratio} shows $\nPt$, i.e., the
signal to noise per mode at $k=0.2\ihMpc$ for the fiducial survey
assumed. The figure shows that $\nPt \lesssim 1$ over the entire redshift
range. \citet{2010ApJ...721..164S} made a conservative decision to assume that BAO
reconstruction is not efficient when a redshift bin has a signal to
noise ratio less than 2. We do not make such assumption in this paper,
but instead directly estimate the reconstruction efficiency at this low
signal-to-noise limit using simulations.

\subsection{BAO reconstruction of the wedged field}

The wedge effect, when not corrected for, would substantially remove the information 
on the underlying density field, thereby degrading an estimation 
of the underlying displacement field as well. Therefore,  the effect of the wedge is not merely a
 cut-off of the contribution from the modes with $\mu<\mumin$ in Eq.~(\ref{eq:windowf}), but should 
also include the degraded efficiency of the BAO reconstruction by using different assumptions of $\Sigxy$ and $\Sigz$ as a function of $\mumin$ after BAO reconstruction.

While the damping scales before the BAO reconstruction are straightforward to calculate analytically using the expected linear theory power spectrum, it becomes nontrivial to predict the corresponding values analytically as a function of noise level after reconstruction, especially in the presence of the wedge effect. \citet{2009PhRvD..79f3523P} has explained the BAO reconstruction technique using the
Lagrangian Perturbation Theory (hereafter ``LPT'') formalism. Using
the formalism, in principle, an analytic approximation of the damping behavior after the BAO reconstruction could be
derived. \citet{2010arXiv1004.0250W} applies the formalism to estimate damping scales in the presence of
various galaxy shot noise levels; however, these analytic calculations
have not been confirmed in detail against the damping scales from simulations and also no redshift-space distortions were considered. Instead of including the wedge effect 
in the analytic formalism, 
in this paper, we directly
estimate the values using \Nbody\ simulations. 

Note that the Gaussian damping function
in Eq.~(\ref{eq:StoN}) is an approximation for the linear information
that survived in the late-time density field, which is often referred to as a ``propagator'' defined in Eq.~(\ref{eq:propa}):  i.e., a cross-correlation
between the late-time density field and the linear density field relative to the linear input power spectrum.
We therefore can estimate $\Sigxy$ and
$\Sigz$ by directly
estimating the propagators as a function of redshift as well as a
function of the noise level. 

\begin{equation}\label{eq:propa}
C(k,\mu,z)=\frac{\langle\hdelta_{\rm lin}(\kvec,z)\hdelta(\kvec',z)\rangle}{\langle\hdelta_{\rm lin}(\kvec,z)\hdelta_{    \rm lin}(\kvec',z)\rangle},
\end{equation}
where $\hdelta_{\rm lin}(\kvec,z)$ is the initial linear fields that is linearly scaled to $z$,     and $\hdelta(\kvec,z)$ is from the final density fields at $z$. 
Then the BAO signal in the nonlinear power spectrum can be modeled as $C(k,\mu,z)^2 \Plin (k,\mu,z)$.

\subsection{\Nbody\ simulations to calculate propagators}
Since the propagator depends on a correlation between the initial field and the final field generated with the same random seed, the sample variance associated with the calculation can be substantially reduced with only one realization\footnote{We also apply Savitzky-Golay smoothing when plotting the propagators to reduce the noise due to the thin $k$-shell averaging. }.
We use one realization of Particle-Mesh dark matter simulations that were used in
\citet{2010ApJ...720.1650S}. The fiducial cosmological parameters were chosen to be
consistent with WMAP5+SN+BAO results \citep{2009ApJS..180..330K}: $\Om=0.279$,
$\OL=0.721$, $h=0.701$, $\Ob=0.0462$, $n_s=0.96$, and
$\sig_8=0.817$. The simulation volume covers $2^3\trihGpc$.
The density field is derived using the Cloud-in-Cell (CIC) method \footnote{In the Cloud-in-Cell method, we decrease the discreteness effect of a mass tracer particle by treating each particle to be a uniform-density cubical cloud with a volume of $(2000/576)^3\trihMpc$ centered at the particle position and count the mass fractions contributed to nearby 8 meshes. } for $576^3$ meshes that cover the volume of $2^3\trihGpc$ and Fourier-transformed.

Note that this cosmology is not identical to the fiducial
cosmology used for the Fisher matrix calculations. We pick two
redshift outputs of one realization, at $z=1$ and $z=2$. We assume
tracers with bias of unity, as shown in Table \ref{tab:fid}, by simply
taking dark matter particles while varying
levels of signal-to-noise by subsampling the dark matter
particles. Although it is possible that, at a fixed bias, propagators for
tracers with different HODs can be slightly different from each other, the difference
has shown to be small and therefore we ignore it \citep[e.g.][]{2009PhRvD..80f3508P, 2009PhRvD..79f3523P, 2011ApJ...734...94M}.

To simulate different values of
$\nPt$,  we
subsample the dark matter particles  at $z=1$ and 2 in the following way. At $z=1$, $\Pt$, i.e., the power at $k=0.2\ihMpc$
is expected to be $716\trihMpc$ using our fiducial cosmology. This requires the
particle number density of $\sim 0.007$, $0.0028$, $0.0014$, $0.0007$,
and $0.00014\itrihMpc$ to mimic $\nPt=5$, 2, 1, 0.5, and 0.1
respectively. The simulation has a particle number density of $0.0239\itrihMpc$, and therefore we
subsample particles by a fraction of 0.29, 0.117, 0.0586, 0.029,
0.00586, respectively. Note that we assume that this Poisson subsampling noise represents the noise 
due to the instrumental and sky background as well as shot noise, $P_{\rm th}({\bmath k})+\Pshot$, in Eq.~(\ref{eq:StoN}). 
We do not explicitly include the effect of the response function to represent a suppression of signal relative to the noise on small scales. The instrument beam is smaller or comparable to the size of the mesh we use for the density calculation and the displacement field calculation, as well as the typical nonlinear scale. Since we expect little information from below the mesh scale, we expect this procedure to accurately capture the distribution of signal-to-noise ratio as a function of ${\bmath k}$.

At $z=2$, linear $\Pt$ is expected to be $342\trihMpc$, which requires the particle number density to be 0.015, 0.0058, and 0.003, 0.0015, $0.0003~\itrihMpc$ to make $\nPt=5$, 2, 1, 0.5, 0.1, respectively. We subsample particles in the simulations by a fraction of 0.628, 0.243, 0.126, 0.0628, and 0.0126, respectively.

\subsubsection{Implementing the wedge}
The minimum $\mu$ angle that defines the wedge effect is approximately
0.55 at $z=1$ and 0.76 at $z=2$ using our fiducial cosmology of the Fisher matrix
analysis and we apply the same value for the simulation despite the small
difference in cosmology. To mimic the wedge effect, we assume that Fourier modes for $\mu<\mumin$ are completely dominated by noise and simply set the corresponding 
Fourier coefficients to be zero. 

The pre-reconstructed power spectrum in the presence of the wedge is straightforward. The squared amplitudes of the Fourier coefficients of the pre-reconstructed density field (after corrected for the CIC pixel-window function) is recorded as a function of $\mu$.

In order to derive a post-reconstructed power spectrum with the wedge effect, we have to first simulate a pre-reconstructed  {\it density field} in the presence of the wedge effect, hereafter $\rhoow$, and  the displacement field that is derived from $\rhoow$. This is done in three stages: (1) the raw pre-reconstructed density field is Fourier transformed; (2) the wedge is applied in Fourier space by zeroing all the resulting Fourier coefficients when $\mu < \mumin$; (3) we then inverse-Fourier transform these wedged Fourier coefficients to construct $\rhoow$. 
Note that the resulting density field is not particle-based any longer, but mesh-based.
The displacement field with the wedge effect is derived in a very similar way, but by modifying the second and the third stages: i.e., (2)  by deriving the wedged $\hat{\bmath q}({\bmath{k}})$ (Eq. ~\ref{eq:qvec}) by zeroing all $\hat{\bmath q}({\bmath{k}})$ when $\mu < \mumin$ and (3) then by an inverse Fourier transformation of $\hat{\bmath q}({\bmath{k}})$. 

We displace the mesh-based $\rhoow$ field according to the derived displacement fields; note that this is different from the conventional way of displacing galaxies, i.e., a particle field.
We find that displacing a mesh-based density field is almost identical to displacing particles as long as the CIC pixel-window function is properly corrected for the mesh-based density field. 
We then displace a uniform density field with the same displacement field and subtract the two in order to derive the final, reconstructed density field $\rhorw$ in the presence of the wedge effect.  

\subsubsection{Smoothing scale}

The BAO reconstruction depends on the smoothing kernel $S(k)$ that
we apply to the nonlinear density field to filter out high $k$ modes when deriving displacement
fields in Eq.~(\ref{eq:qvec}).
The optimal smoothing kernel for BAO reconstruction is shown by \citet{2012JCAP...10..006T}. However,  the goal of the current particlesper is not to
find an optimal BAO reconstruction, but to find an first-order description of the 
qualitative effect of the wedge, and therefore we assume a simple Gaussian smoothing kernel
that has been conventionally used for simulations as well as for data. The
choice of the smoothing length depends on the level of nonlinearity
and also on the level of shot noise. \citet{2010arXiv1004.0250W} shows that using a very small
smoothing length when the high $k$ is noise-dominated  makes
reconstruction ineffective while there exists a broad range of optimal/stable
smoothing length. We want to choose the smoothing length that falls
in this stably optimal range. We choose the
smoothing scale at the noiseless limit to be $5\hMpc$ (i.e., at $\sim$ zero
shot-noise limit) and this corresponds to the smoothing scale of
$5\sqrt{2}\hMpc$ using the convention defined in \citet{2010arXiv1004.0250W}. We would want to alter this scale based on the shot noise level of
subsampled field as shown Eq.~(\ref{eq:Sko}):
\begin{equation}
S({\bmath k}) = e^{-k^2\SigRo^2/2}\frac{1}{1+1/[nP(k)]},\label{eq:Sko}
\end{equation} 
where $1/(1+1/[nP(k)])$ is an optimal weight to filter out high-noise Fourier modes and derive the true $S({\bmath k})\hat{\delta}({\bmath k})$.
Again, we choose $\SigRo = 5\hMpc$.

\begin{figure}
\includegraphics[width=0.8\linewidth]{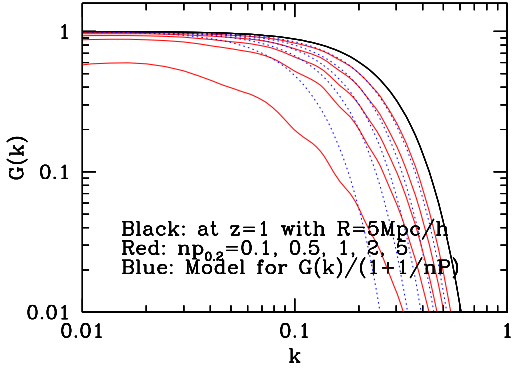}
\caption{The difference in the smoothing kernel model between
  Eq. \ref{eq:Sko} (red) that incorporates the optimal weighting  and Eq.~\ref{eq:Ska} (blue) that we actually use as approximations.} \label{fig:Gvsk}
\end{figure}

For simplicity, however, we approximate this equation by defining an effective
smoothing scale $\SigR$ such that 
\begin{equation}
S({\bmath k}) = e^{-k^2\SigR^2/2},\label{eq:Ska}
\end{equation}
where $\SigR$ is set to produce 
\begin{equation}
e^{-\kt^2\SigR^2/2}= e^{-\kt^2\SigRo^2/2}\,\frac{1}{1+1/[nP(\kt)]}
\end{equation}
at $k=0.2\ihMpc$.
This gives $\SigR=\sqrt{34}$, $\sqrt{45}$, $\sqrt{60}$, $\sqrt{80}$,
and $\sqrt{145}$ $h^{-1}$ Mpc, for $\nPt=5$, 2, and 1, 0.5, and 0.1, respectively.
Figure \ref{fig:Gvsk} shows the difference between Eq. \ref{eq:Sko}
(red) and \ref{eq:Ska} (blue) which quickly becomes worse for low
$\nPt$. We tolerate this difference.

\section{Results}
\label{sec:Results}

In this section, we estimate the propagators at $z=1$ and 2 for various values of $\nPt$ using the \Nb\ simulation and adopt the derived propagators to model the BAO signal in the Fisher matrix predictions for 21cm radio BAO surveys.
\subsection{Reconstruction at $z=1$}\label{subsec:reczo}

\begin{figure*}
  \centering
  \includegraphics[width=1.0\linewidth]{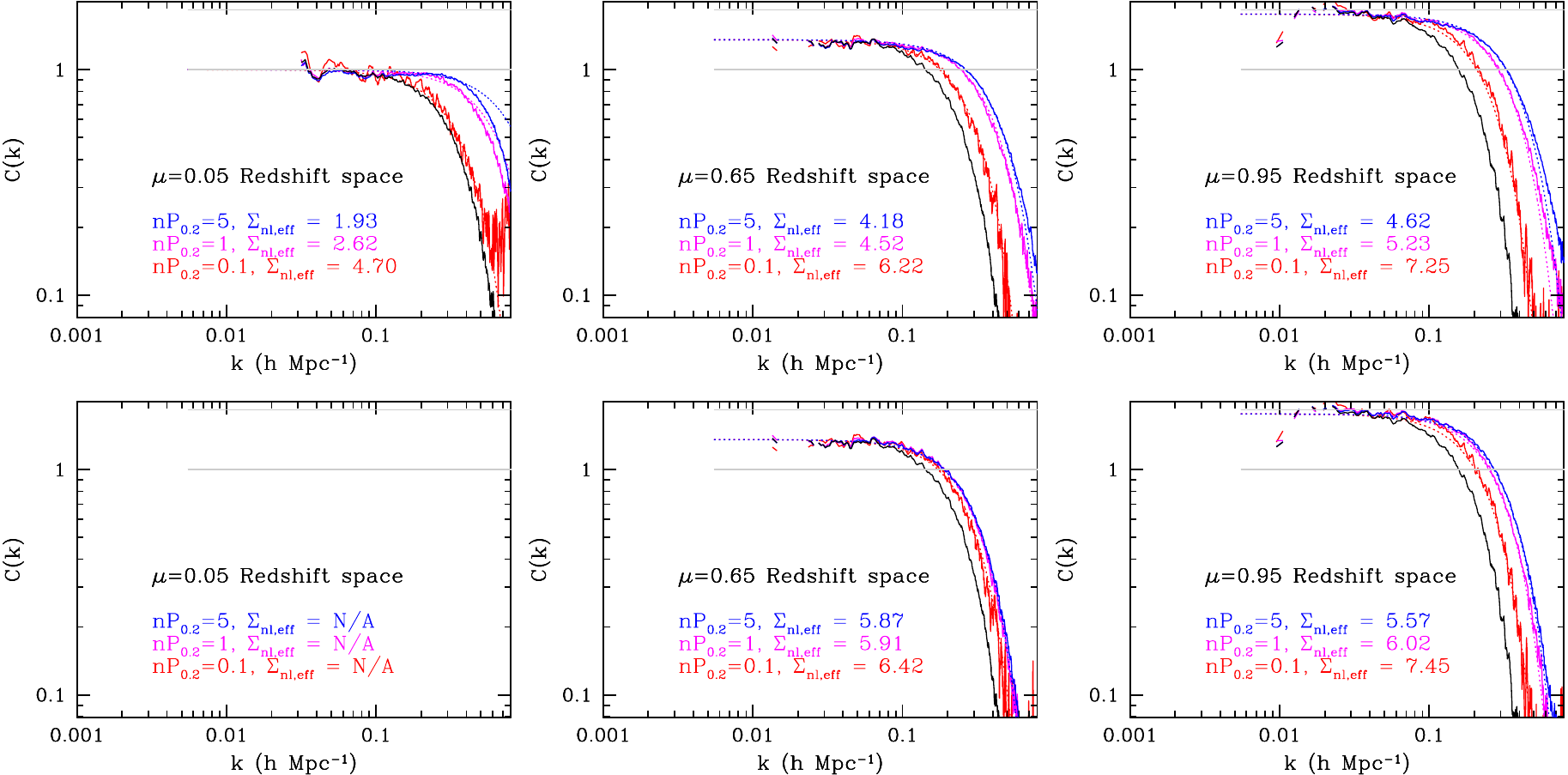}
\caption{Propagators of the subsampled redshift-space matter field, i.e., $\nPt=5$, 1, and 0.1, at $z=1$ of modes in the transverse direction (left), modes along the line of sight (right), and modes just outside of the wedge (middle). Top: without the wedge effect. Bottom: with the wedge effect with $\mumin = 0.55$. Solid lines: measured propagators. Dotted lines: Gaussian model propagators based on the estimated damping scale at given $\mu$ based on $k=0.3\ihMpc$ (i.e., corresponds to the quoted $\Signleff$). Black lines: before reconstruction using the density field with $\nPt=5$. Blue, magenta, and red lines: after reconstruction for $\nPt=5$, 1, and 0.1, respectively. The gray reference lines are at unity and at the linear redshift-space distortion prediction.}\label{fig:zonPf}
\end{figure*}

Figure \ref{fig:zonPf} shows the propagators before (black lines for $\nPt=5$) \footnote{Without BAO reconstruction, propagators for different values of $\nPt$ are very similar and the difference is dominated by the noise.}
and after reconstruction (blue, magenta, and red lines for $\nPt = 5$, 1, 0.1, respectively), without (top) and with this wedge effect (bottom panels). The left panels show the propagators for the 
transverse modes ($\mu=0.05\pm 0.05$), the middle panels show  ones for the modes that are just outside of the wedge ($\mu=0.65\pm 0.05$), and the right panels show ones for the modes along the line of
sight ($\mu=0.95\pm 0.05$). Note that we use modes slightly away from $\mu=0$ and $\mu=1$ 
in order to increase the number of modes and therefore to decrease sample variance in 
estimating $\Sigxy$ and $\Sigz$ from the propagators. The difference in
$\mu^2$ caused by $\Delta \mu = 0.05$ is small, but is accounted for when we derive $\Sigxy$ 
and $\Sigz$.  When estimating $\Sigxy$ and $\Sigz$ for a Gaussian damping model, we use the propagator values at 
$k=0.3\ihMpc$ rather than at $k=0.2\ihMpc$ because, based on our visual inspection, the former seems to give a better description over the scales of
our interest (i.e., $k<0.3\ihMpc$) than the
latter and the larger number of modes at the bin centered at $k= 0.3\ihMpc$ allows the estimation of
the damping length to be more stable. To further decrease the noisy fluctuations in the propagators, we smooth the measured
 propagators using the Savitzky-Golay filtering.   The colored dotted lines correspond to the estimated damping length $\Signleff$ at given $\mu$ based on the propagator value at $k=0.3\ihMpc$.

In the presence of the wedge (bottom panels), the improvement in the modes outside of the wedge (bottom middle and bottom right panels) by the BAO reconstruction (the difference between the colored lines and black lines) is clearly visible, while it is less than in the wedge-less case (the corresponding top panels). The modes in different directions are coupled and the divergence operation is non-local; nevertheless, we find that the linear density field fluctuations along the line of sight is still reconstructable to some extent even in the near absence of the transverse information. 
For example, for $\nPt=5$, we derive $\Sigxy$ and $\Sigz$ to be $5.16\hMpc$ and $10.16\hMpc$, respectively, before reconstruction 
and $1.92\hMpc$ and $4.86\hMpc$ after reconstruction without the wedge, by extrapolating from the measurements of the damping scale at $\mu=0.05$ and $0.95$ to $\mu=0$ and 1. With the wedge effect, $\Sigxy$ and $\Sigz$ are extrapolated from the measurements of the damping scale at $\mu=0.65$ and 0.95; we find $6.12\hMpc$ for $\Sigxy$ and $5.84\hMpc$ for $\Sigz$. The decrease in $\Sigz$ is still substantial even with the wedge effect.
One could consider generating constrained Gaussian realizations to fill the missing information inside the wedge in an attempt to alleviate the effect of the wedge on reconstructing the modes around the wedge \citep{1991ApJ...380L...5H,2012MNRAS.427.2132P}, while our current results show a rather conservative scenario.

The figure shows that the reconstruction efficiency decreases rather slowly when the signal to noise decreases from $\nPt=5$ to 1, but the efficiency decreases more drastically when $\nPt$ decreases from 1 to 0.1. Note that $\nPt=1$ corresponds to the highest signal-to-noise redshift bins in our fiducial survey near $z\sim1$ (Figure \ref{fig:sigDvratio}).

The degradation of the reconstruction efficiency in the presence of the wedge becomes less obvious as the signal-to-noise level $\nPt$ decreases to 0.1, i.e., the largest noise case we investigate. The effect of the wedge on the BAO reconstruction is small in this case. It is probably because the modes within the wedge are already dominated by the noise.

\subsection{Reconstruction at $z=2$}\label{subsec:reczt}

\begin{figure*}
  \centering
  \includegraphics[width=1.0\linewidth]{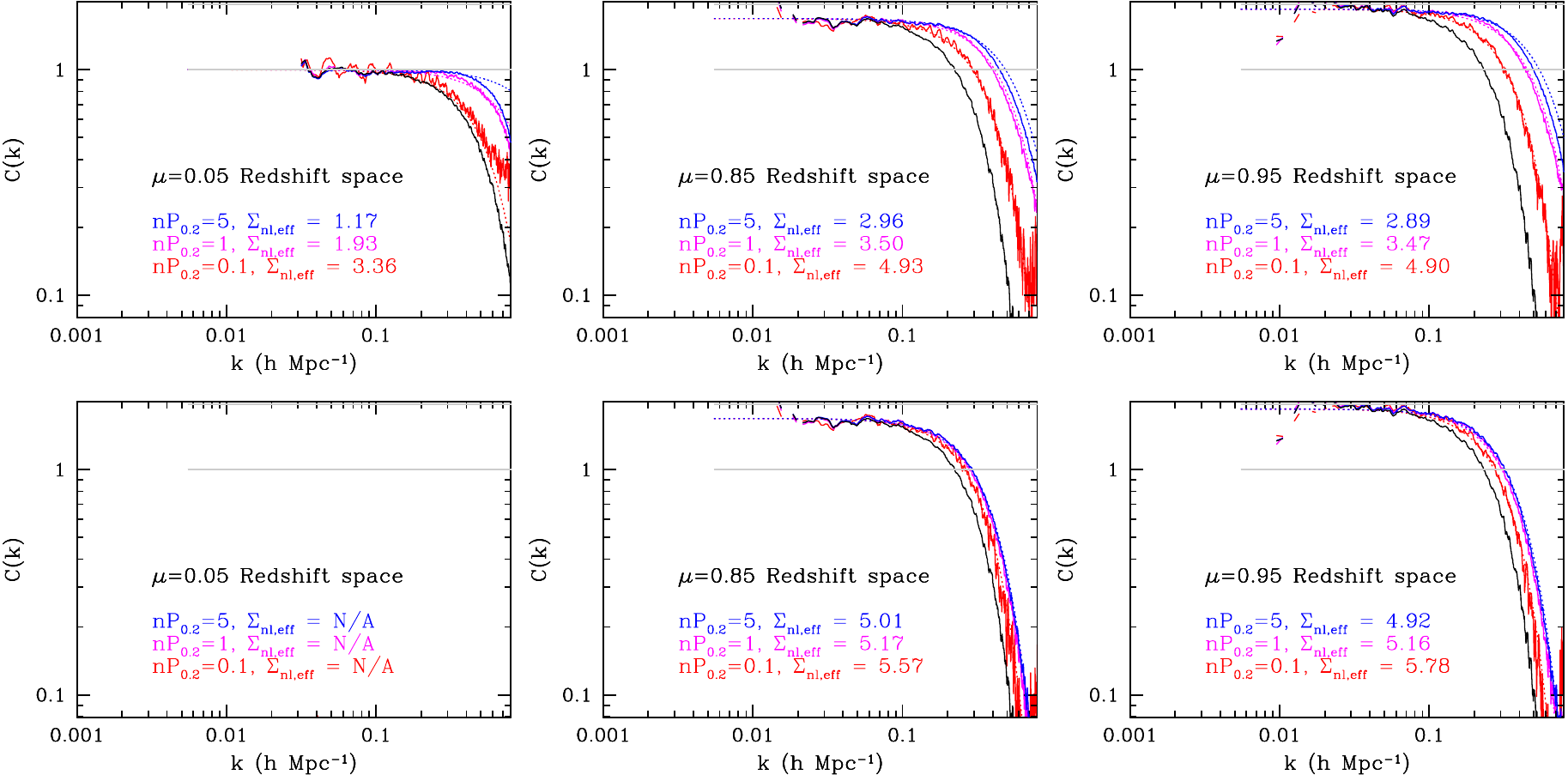}
\caption{Propagators of the subsampled redshift-space matter field, i.e., $\nPt=5$, 1, and 0.1, at $z=2$ of modes in the transverse direction (left), modes along the line of sight (right), and modes just outside of the wedge (middle). Top: without the wedge effect. Bottom: with the wedge effect with $\mumin = 0.76$. Solid lines: measured propagators. Dotted lines: Gaussian model propagators based on the estimated damping scale at given $\mu$ based on $k=0.3\ihMpc$ (i.e., corresponds to the quoted $\Signleff$). Black lines: before reconstruction using the density field with $\nPt=5$. Blue, magenta, and red lines: after reconstruction for $\nPt=5$, 1, and 0.1, respectively. The gray reference lines are at unity and at the linear redshift-space distortion prediction.}\label{fig:ztnPf}
\end{figure*}

A density field at a different redshift would respond differently to the combination of reconstruction and the wedge. We expect the wedge situation to become more serious at higher $z$ since a larger fraction of the modes are lost.
Figure \ref{fig:ztnPf} shows the propagators before
and after reconstruction with and without this wedge effect for the
transverse modes ($\mu=0.05\pm 0.05$), modes that are just outside of the wedge ($\mu=0.85\pm 0.05$, middle panels), and the modes along the line of
sight ($\mu=0.95\pm 0.05$). As expected, the reconstruction efficiency is significanly reduced in the presence of the wide wedge with $\mu \leq 0.76$.
Again $\Sigxy$ and $\Sigz$ are derived using the measurements at $\mu=0.05$ and $0.95$ without the wedge effect, but derived from $\mu=0.85$ and $0.95$ with the wedge effect.
Due to the larger sample variance in the two reference $\mu$ bins of the latter case, the derived estimates of $\Sigxy$ and $\Sigz$ would be more susceptible to noise. We again find that non-zero improvement in the line-of-sight modes outside of the wedge by the BAO reconstruction, but at the level much less than that of the wedge-less case; the degradation due to the wedge is more severe compared to the cases at $z=1$ due to the line-of-sight modes being closer to the wedge boundary. Again, the degradation due to the wedge is less obvious as the noise level increases to $\nPt=0.1$; however, the effect of the wedge appears more severe even at this noise level compared to the same case at $z=1$, probably because more modes are removed. Note that $\nPt=0.1$ corresponds to the lowest signal-to-noise redshift bins in our fiducial survey populated near $z\sim2$ (Figure \ref{fig:sigDvratio}).

\subsection{Propagators as a function of signal-to-noise and the wedge angle}\label{subsec:prop}
Table \ref{tab:tsignl} summarizes the nonlinear damping scales for a
Gaussian approximation we derived, with and without the wedge effect,
and with and without reconstruction at $z=1$ and 2.
 The last column shows the damping scale without reconstruction; it
 was evaluated using a set with $\nPt=5$ since the
  nonlinear degradation before reconstruction in principle would not depend on the noise level $\nPt$.  As a sanity check, we analytically calculated the expected damping
 scale without reconstruction using the method in \citet{2007ApJ...664..660E} and we derived very similar values (within 5 per cent).
 
 \begin{table*}
\centering
\caption{Wedge effect on the non-linear damping scale of BAO. In the case without the wedge effect, in order to reduce the sample variance effect, $\Sigxy$ and $\Sigz$ are extrapolated from the damping scales measured at $0 < \mu  \leq 0.1 $ (centered at $\mu=0.05$)  and $0.9 < \mu < 1$ (centered at $\mu=0.95$), rather than measuring damping scales directly from Fourier modes at $\mu=0$ and 1. In the case with the wedge effect,  $\Sigxy$ and $\Sigz$ are extrapolated from the damping scales measured at $0.6 < \mu  \leq 0.7 $ (centered at $\mu=0.65$)  and $0.9 < \mu < 1$ (centered at $\mu=0.95$) for $z=1$, and from the damping scales at  $0.8 < \mu  \leq 0.9 $ (centered at $\mu=0.85$)  and $0.9 < \mu < 1$ (centered at $\mu=0.95$) at $z=2$.
 Rather than using a formal fitting procedure, we model the propagator as a Gaussian function with $\Sigxy$ and $\Sigz$ evaluated at $k=0.3 \ihMpc$. The damping scale without reconstruction in the last column is evaluated using $\nPt=5$.}
\label{tab:tsignl}
\begin{tabular}{@{}lcccccccccccc}
\hline
 \multicolumn{1}{c|}{}& \multicolumn{10}{c|}{ with reconstruction} &  \multicolumn{2}{c}{ without reconstruction}  \\\hline 
\multicolumn{1}{c|}{} & \multicolumn{2}{c}{ $\nPt = 0.1$} &  \multicolumn{2}{|c}{ $\nPt = 0.5$} &  \multicolumn{2}{|c}{ $\nPt = 1$} &  \multicolumn{2}{|c}{ $\nPt = 2$} &  \multicolumn{2}{|c}{ $\nPt = 5$} & \multicolumn{2}{|c}{ } \\ 
\multicolumn{1}{c|}{} &$\Sigxy$ &  \multicolumn{1}{c|}{$\Sigz$} & $\Sigxy$ & \multicolumn{1}{c|}{$\Sigz$} & $\Sigxy$ & \multicolumn{1}{c|}{$\Sigz$} & $\Sigxy$ & \multicolumn{1}{c|}{$\Sigz$} & $\Sigxy$ & \multicolumn{1}{c|}{$\Sigz$} & $\Sigma_{\rm xy,0}$ & \multicolumn{1}{c}{$\Sigma_{z,0}$} \\\hline
\multicolumn{1}{c|}{$z=1$, $\mumin$ = 0   } 
&  4.69 &\multicolumn{1}{c|}{ 7.63 }  
& 3.09 & \multicolumn{1}{c|}{ 5.96 } 
& 2.61 & \multicolumn{1}{c|}{ 5.51 } 
& 2.28 & \multicolumn{1}{c|}{ 5.18 } 
& 1.92 & \multicolumn{1}{c|}{ 4.86 }
&  5.16 & 10.16 \\ 
 \multicolumn{1}{c|}{$z=1$, $\mumin$ = 0.55} 
& 5.35 & \multicolumn{1}{c|}{ 7.93 }  
& 5.55 & \multicolumn{1}{c|}{ 6.72 }  
& 5.81 &  \multicolumn{1}{c|}{6.35 }  
& 5.97 & \multicolumn{1}{c|}{ 6.10 }  
& 6.12 & \multicolumn{1}{c|}{ 5.84 } 
&     & \\\hline
 \multicolumn{1}{c|}{$z=2$, $\mumin$ = 0  } 
& 3.36 & \multicolumn{1}{c|}{ 5.16 }
& 2.17 & \multicolumn{1}{c|}{ 3.99 }
& 1.92&  \multicolumn{1}{c|}{ 3.65 }
& 1.53 & \multicolumn{1}{c|}{ 3.29 }
& 1.16 & \multicolumn{1}{c|}{ 3.04 }
&     3.59 & 7.26 \\
\multicolumn{1}{c|}{$z=2$, $\mumin$ = 0.76 }
& 4.63 &\multicolumn{1}{c|}{ 6.17 }
& 5.44 &\multicolumn{1}{c|}{ 5.56 }
& 5.21 &\multicolumn{1}{c|}{ 5.43 }
& 5.40 &\multicolumn{1}{c|}{ 5.24 }
& 5.36 &\multicolumn{1}{c|}{ 5.14 }
& &\\
\hline
\end{tabular}
\end{table*}

\begin{figure*}
\begin{center}
  \includegraphics[width=0.9\linewidth]{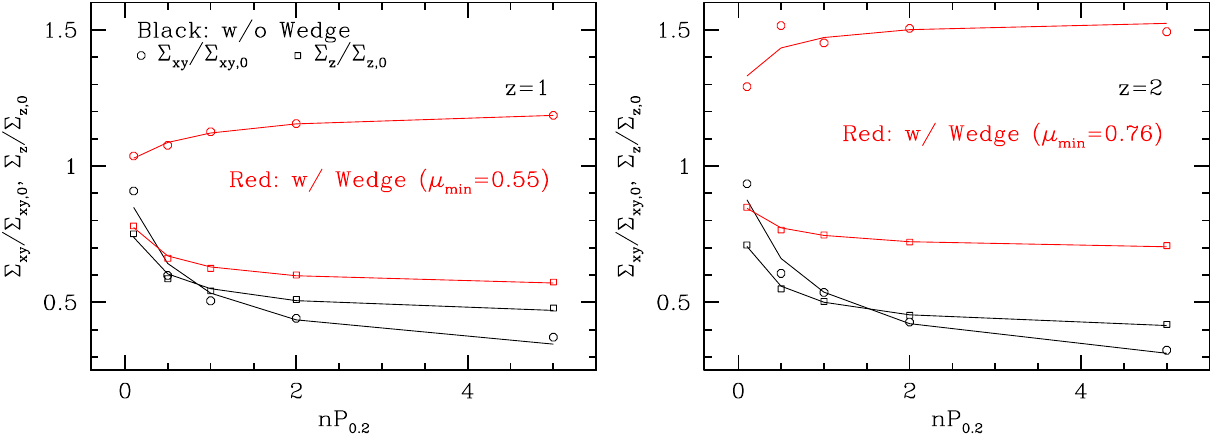}
\caption{The effect of the wedge on reconstructed nonlinear damping scales from Table \ref{tab:tsignl}. The damping scales are shown as ratios with respect to the values without reconstruction, i.e., $\Sigxyo$ and $\Sigzo$.  A smaller ratio below unity indicates more efficient BAO reconstruction. The black points show the results without the wedge effect and the red points show the results with the wedge effect. Circles show $\Sigxy/\Sigxyo$, i.e., the damping scale ratio for the transverse modes and squares show $\Sigz/\Sigzo$, i.e., the damping scale ratio for the line-of-sight modes.
The lines are our models for the data points, i.e., $1-a_0(\nPt/(1+\nPt))^{a_1}$, and the best fit $a_0$ and $a_1$ in each case are described in the main text.}\label{fig:fsignl}
\end{center}
\end{figure*}

Figure \ref{fig:fsignl} visualizes the results in Table
\ref{tab:tsignl}, i.e.,  the improvement in the nonlinear damping scale
due to reconstruction as a function of $\nPt$ in the presence of the
wedge effect (red) and no wedge effect (black); the damping scales are
shown as ratios with respect to the values without reconstruction,
i.e., $\Sigxyo$ and $\Sigzo$, and therefore a smaller value
corresponds to a bigger improvement. We find that the dependence of the reconstruction efficiency on $\nPt$ is moderate from $\nPt=5$ to $\nPt=0.5$ \citep[also see][]{2011ApJ...734...94M} while it is substantial when approaching $\nPt=0.1$. That is, we do not seem to gain much in terms of reconstruction by improving $\nPt >2$. 

From the left panel of Figure \ref{fig:fsignl}, the smaller difference between the red and the black squares visualizes our earlier finding that the reconstruction efficiency for the modes along the lines of sight is not greatly affected by the presence of the wedge at $z=1$, while the large difference between the red and black circles shows that the efficiency degradation when moving away from the line of sight is very large. The right panel visualizes the result at $z=2$, i.e., a much larger degradation in the reconstruction efficiency due to the presence of the larger wedge. The differences between the red and black points decrease as $\nPt$ decreases to 0.1, which again shows that the effect of the wedge is less as the noise level increases. 
The solid lines are fits to the ratios of the measured nonlinear damping scales assuming a fitting function in a form of
\begin{equation}
\frac{\Sigma_i}{\Sigma_{i,0}} = 1-a_0\left(\frac{\nPt}{1+\nPt}\right)^{a_1}.
\label{eq:a0a1}
\end{equation}
Here $i=xy$ or $z$ is the direction (transverse or radial) along which we consider nonlinear smearing. The parameter $a_0$ describes the fractional improvement in the nonlinear smearing in the limit of high signal-to-noise ratio ($\nPt\gg1$), whereas $a_1$ describes how that improvement depends on $\nPt$ ($a_1=1$ would mean that half of the improvement is realized at $\nPt=1$, whereas $a_1<1$ means that more than half of the improvement is realized at $\nPt=1$). The fit parameters are shown in Table~\ref{tab:fitpars}.

\begin{table}
\caption{\label{tab:fitpars}The fit parameters $a_0$ and $a_1$ in Eq.~(\ref{eq:a0a1}).}
\begin{tabular}{ccccccc}
\hline\hline
Case & & \multicolumn2c{$\Sigma_{xy}$} & & \multicolumn2c{$\Sigma_z$} \\
 & & $a_0$ & $a_1$ & & $a_0$ & $a_1$ \\
\hline
$z=1$, no wedge & & $+0.737$ & $0.659$ & & $+0.562$ & $0.319$ \\
$z=1$, wedge & & $-0.217$ & $0.837$ & & $+0.453$ & $0.290$ \\
$z=2$, no wedge & & $+0.792$ & $0.773$ & & $+0.619$ & $0.309$ \\
$z=2$, wedge & & $-0.544$ & $0.207$ & & $+0.312$ & $0.291$ \\
\hline\hline
\end{tabular}
\end{table}

\begin{figure*}
\centering
  \includegraphics[width=0.8\linewidth]{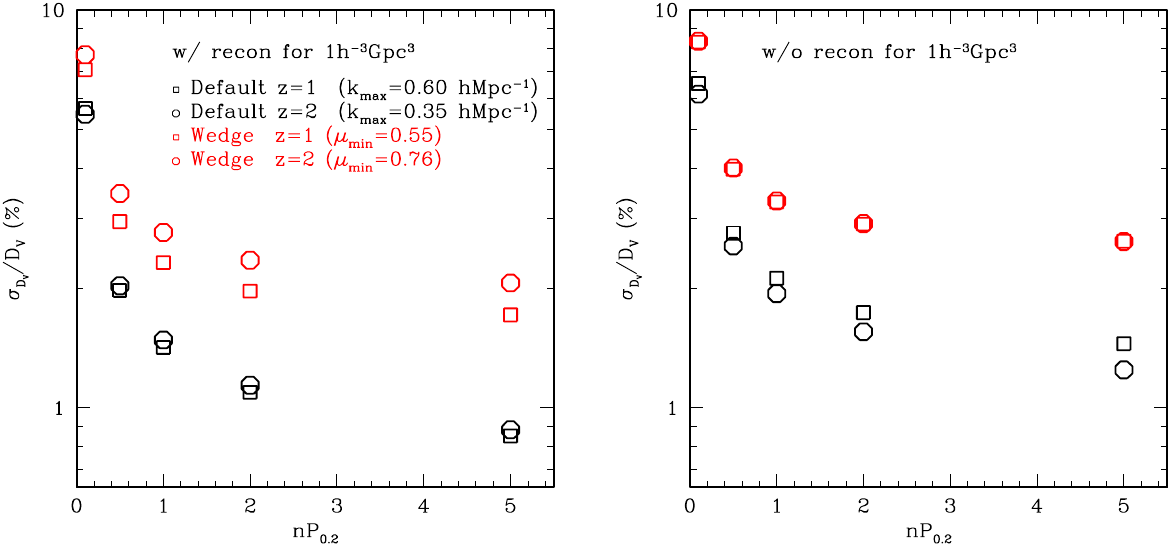}
\caption{The effect of wedge on the isotropic distance precision as a function of $\nPt$. We assume a volume of 1 $\trihGpc$ rather than the actual volume of each redshift bin of the fiducial survey. We assume the transverse resolution limit for the fiducial survey applies here such that the maximum $\kmax = 0.6 \ihMpc$ at $z=1$ (squares) and 0.35 at $z=2$ (circles). Left: with reconstruction. Right: without reconstruction. The red points show the results with the wedge effect and the black points show the results without the wedge effect.  }\label{fig:fsigDV}
\end{figure*}

For the radial nonlinear smearing length $\Sigz$, the main effect of the wedge is a modest reduction in $a_0$. That is, if the wedge is excised, the reconstruction algorithm performs with the same dependence on $\nPt$ ($a_1\approx 0.3$) but the amount of improvement ($a_0$) is reduced by 20 per cent at $z=1$ and 50 per cent at $z=2$.

The transverse nonlinear smearing length $\Sigxy$ exhibits more complex behavior. When there is no wedge effect, the dependence of $\Sigxy/\Sigxyo$ on $\nPt$ is stronger than that of $\Sigz/\Sigzo$ (larger $a_1$), such that $\Sigxy/\Sigxyo$ is smaller than $\Sigz/\Sigzo$ at high values of $\nPt$. This makes sense because the redshift-space distortions increase the effective $nP({\bmath k})$ for the more radial modes that lead to displacements in the $z$-direction.
The fact that $\Sigxy/\Sigxyo$ is larger than unity for the wedged case ($a_0<0$; see also the red circles in Fig.~\ref{fig:fsigDV}) appears at first sight to suggest that the reconstruction has made the nonlinearity problem worse in the transverse direction. However, we should recall that for a general Fourier mode with direction cosine $\mu$, the effective nonlinearity scale is neither the transverse scale $\Sigxy$ nor the radial scale $\Sigz$, but rather the combination $\sqrt{\Sigxy^2(1-\mu^2) + \Sigz^2\mu^2}$. That is, $\Sigxy$ is merely a parametric presentation of the damping scale for the modes in the ``dirty wedge'' after reconstruction, extrapolated to $\mu=0$. The actual damping scale in the ``clean window'' (large $\mu$) decreased after reconstruction, as shown in Figures \ref{fig:zonPf} and \ref{fig:ztnPf}.

\subsection{The effect of the foreground on small line-of-sight wave modes during the BAO reconstruction}
We have mentioned that the first order impact of the spectrally smooth astrophysical foreground would be losing the very small line-of-sight Fourier modes due to its contamination, and thereby we focused on the second order impact such as the foreground wedge that is more extensive in Fourier space.  
Again, the removal of the small line-of-sight modes has little effect on the BAO Fisher error forecasts due to the small number of modes, and therefore we have ignored these $k_\parallel$ cutoff and assumed that the fluctuation in foreground contributes the additional noise that corresponds to 10K across all Fourier modes.
However, the statement above warrants some caution when we consider the BAO reconstruction. While there are not many modes on large scales,  the differential motions such as the displacement fields are substantially contributed by large-scale modes below $k \lesssim 0.1\ihMpc$ and even below  $k \lesssim 0.04\ihMpc$ \citep[e.g.,][]{2007ApJ...664..675E}.  In this subsection, we test the effect of ignoring the low $k_\parallel$ cutoff in deriving the displacement fields.  

\citet{2008PhRvL.100i1303C} presents very conservative limits for the smooth foreground subtraction, in terms of the wave numbers where the residual of the foreground subtraction due to the uncertainty in the foreground modeling exceeds the cosmological signal. Their estimate gives $k \sim 0.01\hMpc$ for $z\sim 1$ and $k\sim 0.02\hMpc$ at $z=2$.  As \citet{2010ApJ...721..164S} argues, this is a quite pessimistic limit. \citet{2012MNRAS.419.3491L}, although they focused on the Epoch of reionization, shows that the foreground is featureless enough that it can be characterized to a great accuracy using only three or four independent parameters. Assuming that we need four Fourier modes to determine four independent parameters for the foreground  at each transverse location, we can derive $k_{z,\rm min} = 4\times \pi/\Delta r$, where $\Delta r$ is the radial range of the survey: for $N$ polynomials, we would need  $k_{z,\rm min} = (N+1) \times \pi/\Delta r$. 

For the fiducial survey in this paper, $\Delta r \sim 2000\hMpc$ between $0.7<z<2.11$, which is equivalent to 0.60 e-folds in frequency.  We choose $k_{z,\rm min}=0.01\ihMpc$ that corresponds to six ($=N+1$) free parameters. Since our simulation box has the same dimension as $\Delta r$, this corresponds to removing six line-of-sight Fourier modes in the simulation box as well.   When we remove all the modes for $k_{z,\rm min} \leq 0.01\ihMpc$ at $z=1$ from the density field (and therefore from the displacement field) in the case of $\nPt=5$, i.e., the most effective reconstruction case, we find that the line-of-sight damping after reconstruction, $\Sigz$, increases by 16 per cent, such that $\Sigz/\Sigzo$ in Figure \ref{fig:fsignl} increases from 0.57 to 0.66. Meanwhile, there was little impact on $\Sigxy$, which makes sense since the wedge has already removed most of the information in the transverse direction that would be removed by $k_{z,\rm min}$.  In the lowest signal-to-noise case, i.e., $\nPt=0.1$, we find a smaller difference: an increase of 7 per cent in $\Sigz$. With $k_{z,\rm min}=0.007\ihMpc$, which corresponds to four free parameters, we find 11 per cent increase in $\Sigz$ with $\nPt=5$. With $k_{z,\rm min}=0.007\ihMpc$, which corresponds to four free parameters, we find 11 per cent increase in $\Sigz$ with $\nPt=5$.

At $z=2$, removing $k_{z,\rm min} \leq 0.01\ihMpc$  gives $\Sigz = 5.47\hMpc$ after reconstruction for $\nPt=5$, which is only 6 per cent larger than the value in Table \ref{tab:tsignl}. This value will increase $\Sigz/\Sigzo$ for $\nPt = 5$ in Figure \ref{fig:fsignl} from 0.71 to 0.75.   The difference in $\Sigz$ remains at 7 per cent at $\nPt=0.1$. Taking $k_{z,\rm min} \leq 0.02\ihMpc$ would increase $\Sigz$ after reconstruction (for $\nPt=5$) by 13 per cent. 

We therefore conclude that our assumption of ignoring the loss of the very small line-of-sight Fourier modes due to the smooth foreground component is fairly good approximation even when considering the BAO reconstruction.

\subsection{Effect of the wedge on the distance measurements}

We next incorporate the resulting propagator estimates into the Fisher matrix calculation and estimate distance precisions while assuming our fiducial telescope configuration but only at redshift bins located at $z=1$ and 2. In detail, we assume $\nPt=0.1, 0.5, 1.0, 2, 5.0$ and a survey volume of $1\trihGpc$ in Eq.~(\ref{eq:StoN}) while taking a fiducial power spectrum and the resolution limit $\kmax$ for $\hat{W}^2$ at the corresponding redshifts defined by our fiducial configuration. Nonlinear damping scales $\Sigxy$ and $\Sigz$ are taken from Table \ref{tab:tsignl} given the redshift and the noise level.

\begin{figure*}
 \centering
  \includegraphics[width=0.9\linewidth]{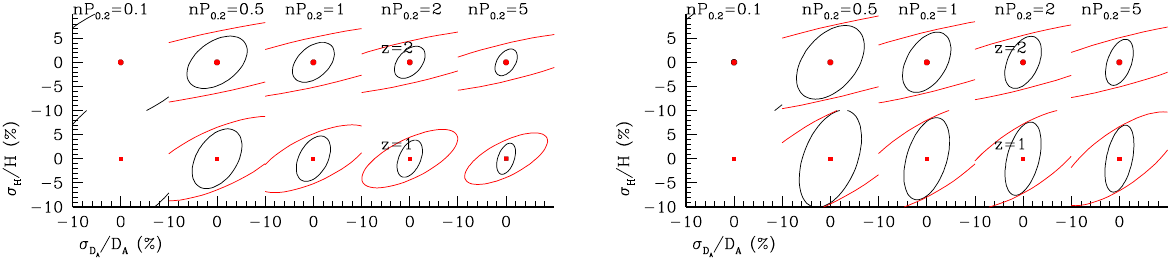}
\caption{The effect of wedge on angular diameter distances and hubble parameters as a function of $\nPt$ . We assume a volume of 1 $\trihGpc$  rather than the actual volume of each redshift bin of the fiducial survey. We assume the resolution limit for the fiducial survey applies here such that the maximum $\kmax = 0.6 \ihMpc$ at $z=1$ (squares) and 0.35 $\ihMpc$ at $z=2$ (circles). Left: with reconstruction. Right: without reconstruction. The red ellipses show the results with the wedge effect and the black ellipses show the results without the wedge effect. One sees that the wedge effect is more severe if we consider BAO reconstruction. }\label{fig:fsigDAHZ}
\end{figure*}

From the right panel of Figure \ref{fig:fsigDV}, i.e., without reconstruction, one can see that the distance precisions at $z=2$ is slightly better than those at $z=1$ without the wedge effect.
If the effect of the response function at the two different redshifts (i.e., $\kmax$ in $\hat{W}^2$) were ignored, the distance precision at $z=2$ would have been noticeably better due to the smaller nonlinear damping effect. With $\kmax=0.35\ihMpc$ that limits $\hat{W}^2$ along the transverse direction at $z=2$, compared to $0.6\ihMpc$ at $z=1$, the precision of $\DA$ at $z=2$ becomes substantially deteriorated, making the overall precision of $\DV$ only slightly better at $z=2$ with help from a larger value of $\beta$.  The $1-\sigma$ error contours of $\DA$ and $\Hz$ in the right panel of Figure \ref{fig:fsigDAHZ} (black ellipses) shows that the precision of $\DA$ constraints is slightly worse at $z=2$ while the precision of $\hz$ is much better at $z=2$. 

With the wedge effect (red points in the same panel), the error on $\DV$ increases by a factor of 1.27--1.8 (i.e., larger at higher $\nPt$) at $z=1$ and 1.36--2.1 at $z=2$ reflecting the more severe wedge effect at $z=2$. As a result, the error on $\DV$ with the wedge effect are coincidently almost the same at $z=1$ and $z=2$. In the right panel of Figure \ref{fig:fsigDAHZ} (red error ellipses), one can see that $\DA$ is virtually not constrained at the $z=2$ bin.

The left panel of Figure \ref{fig:fsigDV}, shown in comparison to the right panel, shows that the reconstruction is more effective at $z=1$ except at the very low signal-to-noise limit. Figure \ref{fig:fsignl} implies a slightly more effective reconstruction at $z=2$ in terms of the damping scale decreasing, but decreasing damping scale brings a larger return at the lower redshift when the field is more nonlinear to begin with.
By comparing the red points in the two panels, we calculate the reconstruction efficiency in the presence of the wedge; the fractional decrease in error is 0.85--0.65 at $z=1$ and 0.93--0.79 at $z=2$. Therefore, despite the wedge effect, the survey would benefit substantially from the BAO reconstruction.

 The left panel of Figure \ref{fig:fsigDAHZ} shows the $1-\sigma$ ellipses of $\DA$ and $\hz$ with reconstruction. Note that the ellipses with the wedge effect (red) do not touch ellipses without the wedge effect (black) after reconstruction, which is distinguished from the cases (the right panel) without the reconstruction. This is reasonable in that, with reconstruction, the wedge effect does more than just wiping out the first-hand transverse information; in the presence of the wedge, the reconstruction of modes along the lines of sight is also affected by the missing transverse modes because of the worse fidelity of displacement fields.  On the other hand, we find that if we do not remove the transverse modes until after we derive displacement fields (i.e., when the displacement fields are correctly derived and the wedge is then applied),  the ellipses of the two cases (red and black in the left panel) tend to touch each other at least at $z=1$. 

Finally, we apply the propagator estimates to all the redshift bins of our fiducial survey which have different fiducial power spectrum amplitudes, resolution limits, $\nPt$, and the survey volumes at different redshift bins that are included.
The fiducial 21 cm BAO survey we assumed has $\nPt \sim 1$ at $z\sim 0.7$ and decreases to $\sim 0.1$ at $z\sim 2$ (Figure \ref{fig:sigDvratio}).
However, Figure \ref{fig:fsignl} shows a non-negligible level of reconstruction efficiency even at $\nPt \sim 1$. We calculate the nonlinear damping scales for the WMAP1 cosmology by taking $\Sigxyo=9.38\hMpc $\footnote{The value is rather large because of a large $\sig_8(=0.9)$ for this cosmology.} at $z=0$, rescale $\Sigxyo(z)$ with the growth function, and derive $\Sigz$ accounting for redshift-space distortions. The reconstruction is incorporated by making a use of the fitting models for $\Sigxy/\Sigxyo$ and $\Sigz/\Sigzo$, i.e., $1-a_0(\nPt/(1+\nPt))^{a_1}$ with the estimates of $a_0$ and $a_1$ from \S~\ref{subsec:prop}. Since we have propagator estimates for only two redshifts, i.e., at $z=1$ and 2, we apply $a_0$ and $a_1$ derived at $z=1$ to the redshift bins between 0.7 and 1.5, and $a_0$ and $a_1$ from $z=2$ to  the redshift bins between 1.5 and 2. 

\begin{figure*}
 \centering
  \includegraphics[width=0.9\linewidth]{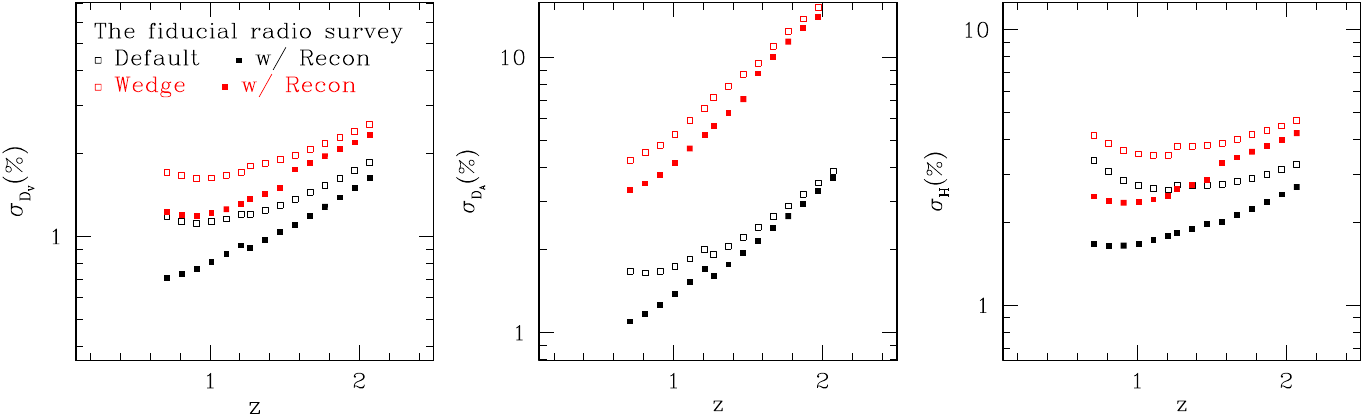}
\caption{The effect of the wedge on distance constraints predicted for the fiducial survey. The open symbols show distance constraints without the BAO reconstruction, and the sold symbols include the BAO reconstruction. Black points: default, without the wedge effect. Red points: with the wedge effect.}\label{fig:fsigfid}
\end{figure*}

Focusing on $\hz$ constraints (the right panel of Figure \ref{fig:fsigfid}), we find that, the reconstruction improved the precisions from the wedged field (compare the red solid and red open points) by $40$ per cent near $z\sim 1$ and  $10$ per cent at $z\sim 2$. For the readers who are interested in comparisons to \citet{2010ApJ...721..164S}, near $z\sim 1$, the reconstructed field with the wedge effect (solid red) would give a better precision on $\hz$ than the un-reconstructed field without the wedge effect (open black, i.e., similar to the default from \citet{2010ApJ...721..164S} except for the assumption of the response function), while the reconstruction helps little near $z\sim 2$ due to the very low $\nPt$ of the reference survey. For $\DA$ constraints (middle), as expected, the wedge effect is detrimental; nevertheless, the reconstruction makes $\sim 20$\% improvement in errors on $\DA$ near $z\sim 1$ despite the wedge effect. The left panel shows the resulting constraints on $\DV$. 

In order to estimate the overall effect of the wedge when BAO reconstruction is applied, one would want to compare the difference between the solid black points and the solid red points; the wedge effect increases the error on $\hz$ by a factor of 1.5-1.6, the error on $\DA$ by 3-4.4 times. In addition, the wedge effect substantially increases the cross-correlation coefficient from the original 0.41 (low $z$)--0.44 (high $z$) to 0.62--0.74, reflecting the increased degeneracy due the poorly constrained $\DA$. Reconstruction slightly increases the coefficient for the wedged case, probably due to a larger offset between the error on $\DA$ and $\hz$ after reconstruction.

\section{Conclusions}
\label{sec:Con}
Foregrounds from Galactic and extra-Galactic origins in combination with the chromatic response of the instruments are expected to contaminate a larger wedged volume in Fourier space around the transverse modes. In this paper, we investigated the effect of the contaminated wedge on the future 21cm radio BAO surveys considering the BAO reconstruction technique.  

We first used \Nb\ simulations to find the efficiency of the BAO reconstruction in the presence of the wedge and compared it to the wedge-less case for various noise levels. We found that the BAO reconstruction can still improve the BAO signal for the Fourier modes along the line of sight despite the near absence of the transverse information. The efficiency is sightly less in comparison to the wedge-less case, and the degradation increases quickly for modes away from the line of sight.  Such degradation also increases as the redshift increases due to the larger wedge at higher redshift. The degradation becomes less obvious as the density field becomes noisier. 

We find that the dependence of the reconstruction efficiency on the signal to the noise level per mode, as parameterized with $\nPt$, is moderate from $\nPt=5$ to $\nPt=0.5$ while it is substantially worse when approaching $\nPt=0.1$. The ability to do reconstruction saturates at high $\nPt$, with little improvement at $\nPt >2$. The fiducial 21 cm BAO survey considered here operates in a low signal-to-noise regime with $1>\nPt>0.1$ at $1<z<2$. The BAO reconstruction appears still valuable for this level of noise. We derive fitting formula for the BAO damping scales as a function of noise with and without the wedge effect. 

The overall effect of the wedge for our fiducial 21 cm radio survey was derived using the Fisher matrix analysis utilizing our prediction on the BAO reconstruction efficiency based on the findings written above. As expected, the effect of the wedge is detrimental to the measurements of the angular diameter distance; the errors on angular diameter distances increased by 3-4.4 times.  The wedge effect increases the error on $\hz$ by a factor of 1.5-1.65 on the other hand.  Again, the reconstruction was still effective even with the wedge. For $\hz$, we observed 10--40 per cent of a precision improvement between $z=1 \;(\nPt =1, \mumin=0.55)$ and $2 \;(\nPt=0.1, \mumin=0.76)$, which was up to 10 per cent less improvement compared to the wedge-less case. Even for $\DA$, we still see 20 per cent improvement despite the wedge.

In summary, we find that the calibration technique to lessen the wedge effect would be extremely valuable for obtaining compelling measurements of angular diameter distances from the 21 cm BAO surveys with the advent of near-future BAO surveys in other wavelengths. For example, \citet{2014ApJ...781...57S,2015PhRvD..91h3514S} show that given the perfect knowledge of the telescope we can clean the foregrounds even with polarization well within the wedge down to $\kpar \sim 0.02\hMpc$, which essentially would be the same as the wedgeless case.

\section*{Acknowledgements}
We are thankful to John Marriner for very useful conversations on the response function and to Miguel Morales, Ue-Li Pen, Martin White, Daniel J. Eisenstein, Philip Bull, and Kris Sigurdson for very valuable comments.
C.H. is supported by the David and Lucile Packard Foundation, the Simons Foundation, and the U.S. Department of Energy. 
H.-J.S was partly supported by Center for Cosmology and Astroparticle Physics, Ohio State University.

\end{document}